\documentstyle[prb,preprint,aps,epsf,amstex,amssymb]{revtex}
\begin{document}
\title{Electron Correlations in an Electron Bilayer at Finite
Temperature: Landau Damping of the Acoustic Plasmon} 
\author{ D.~S.
Kainth,$^{*}$ D. Richards,$^{*}$ H.~P. Hughes,$^{*}$ M.~Y.
Simmons,$^{\dagger}$ and D.~A. Ritchie$^{\dagger}$}
\address{*~Optoelectronics Group; $\dagger$~Semiconductor Physics
Group} 
\address{Cavendish Laboratory, Madingley Road, Cambridge CB3
OHE, U.K.} 
\date{\today} 
\maketitle
\begin{abstract}
We report angle-resolved Raman scattering observations of the
temperature dependent Landau damping of the acoustic plasmon in an
electron bilayer system realised in a GaAs double quantum well
structure.  Corresponding calculations of the charge-density
excitation spectrum of the electron bilayer using forms of the random
phase approximation (RPA), and the static local field formalism of
Singwi, Tosi, Land and Sj\"{o}lander (STLS) extended to incorporate
non-zero electron temperature $T_{\rm e}$ and phenomenological
damping, are also presented.  The STLS calculations include details of
the temperature dependence of the intra- and inter-layer local field
factors and pair-correlation functions.  Good agreement between
experiment and the various theories is obtained for the acoustic
plasmon energy and damping for $T_{\rm e} \lesssim T_{\rm F}/2$, the
Fermi temperature. However, contrary to current expectations, all of 
the calculations
  show significant departures from our experimental
  data for $T_{\rm e} \gtrsim T_{\rm F}/2$. From this, we go on to 
  demonstrate unambiguously that real local field factors fail to provide
  a physically accurate description of exchange correlation
  behaviour in low dimensional electron gases. Our results suggest 
  instead that one must resort to a {\em{dynamical}} local field theory,
  characterised by a {\em{complex}} field factor to provide a more accurate
  description.
\end{abstract}

\section{Introduction}

Low-dimensional electron gas systems confined in semiconductor
heterostructures provide ideal systems for the study of
electron-electron interactions, providing a high degree of structural
quality and control.  The usual theoretical approach adopted for the
description of the collective excitation spectra of such systems is
the Random Phase Approximation (RPA).  However, the RPA does not
include the effects of exchange and correlation, which are expected to
be much more important in two-dimensional electron gas (2DEG) systems
than in 3D, and to increase in importance as the number density of the
system decreases.\cite{Jonson76} To account for these effects, many
authors have gone beyond the RPA for a description of the excitation
spectra of such systems,\cite{Jonson76,Gold92,Neilson93,Zheng94,Liu96}
by adopting the local field theories of Hubbard\cite{Hubbard57} and
Singwi, Tosi, Land and Sj\"{o}lander (STLS),\cite{Singwi68} which
include corrections for exchange and correlation in a simple,
physically motivated manner;\cite{STLS,Mahan} the success of these
approaches can be seen by comparing their results against Monte Carlo
calculations.\cite{Monte Carlo} There have also been reports of a
number of alternative approaches for including exchange-correlation
effects in the excitation spectra of two-dimensional electron systems
by adopting: (i) a non-local approach within a Hartree-Fock
approximation for the evaluation of the irreducible polarisability
(see, \textit{e.g.}, Refs.  \onlinecite{Ryan} and \onlinecite{Luo93}
for the case of intersubband excitations in quantum wells); (ii)
alternative static local field approaches such as the quasi-localized
charge method;\cite{Kalman} (iii) a dynamic local field approach
within the quantum STLS theory.\cite{Mougdil95} However, the static
local field STLS approach has remained a firm favourite in the
published literature.  Tests of local field theories for
three-dimensional systems have been reported for, \textit{e.g.},
Aluminium\cite{Larson98} but there are relatively few reliable
experimental tests of these widely used theories in the density range
applicable to semiconductor heterostructures. We report here such 
an experimental test and demonstrate that
static local field corrections are unable to 
provide an accurate 
description of exchange correlation behaviour in low 
dimensional electron gases. Indeed we will argue that our results
point to the need to examine {\em{dynamical}} local field theories, embodied
by a {\em{complex}} local field factor to achieve a more accurate
description of our experimental results.

Electron bilayers (two parallel 2DEGs in close proximity, produced by
MBE-growth of a GaAs/AlGaAs modulation-doped double quantum well
heterostructure) have been identified as useful systems for the study
of electron correlations in low-dimensional electron systems because
the inter-layer Coulomb interaction can effectively counterbalance the
kinetic energy of the electrons, allowing many body effects to
dominate.  Such a system supports two plasmon modes corresponding to
the in-phase (optic plasmon, OP) and out-of-phase (acoustic plasmon,
AP) intra-sub-band oscillations of the charge densities in each
layer.\cite{DasSarma81} Liu \textit{et al.} have shown that electron
correlations can push the acoustic plasmon curve completely into the
single-particle excitation (SPE) spectrum, resulting in Landau damping
of the mode at much smaller wavevectors than predicted by the
RPA.\cite{Liu96} The acoustic plasmon may therefore be used as a probe
of exchange-correlation effects in 2DEG systems, and of Landau damping
effects.

Raman spectroscopy is a powerful tool for the study of the electronic
excitation spectra of low-dimensional semiconductor
systems.\cite{LSSIV,LSSV} For example, measurements of intersubband
excitations in quantum wells have indicated the importance of
exchange-correlation in these systems.\cite{Gammon92} Applying angle
resolved Raman techniques to electron bilayers, we have recently
determined the dispersion of the acoustic plasmon of a double quantum
well structure.\cite{Kainth98,Kainth99} However, the in-plane
wavevectors $q$ accessible ($q \lesssim 0.15\ k_{\rm F}$ [the Fermi
wavevector], for 2DEG areal densities $N_{\rm s}\sim 2 \times 10^{11}
{\rm cm}^{-2}$), determined by the wavelength of the incident light,
are too small to probe exchange-correlation corrections to the AP
dispersion, as predicted in Ref.  \onlinecite{Liu96}.  Attempts to
extend the wavevector range by overlaying a grating to couple in
higher wavevector transfers\cite{Zettler89} did not provide
significantly clear results for any conclusions to be drawn; the
superposition of the folded AP and OP branches in the reduced
Brillouin zone of the grating leads to ambiguities in any analysis of
the form of the plasmon dispersions.\cite{Brad} However, substantial
overlap between the SPE continuum and the AP can be obtained by
increasing the electron temperature, resulting in the Landau damping
of the plasmon, as we have demonstrated recently.\cite{Kainth98}

In this earlier work we characterised the dependence of the Landau
damping of the AP mode in a bilayer system on electron temperature
$T_{\rm e}$ by comparing the AP peak widths measured using Raman
spectroscopy with those calculated using the RPA and the RPA including
a Hubbard (static, zero-$T_{\rm e}$) local field correction (H-RPA) to
account for exchange-correlation effects.\cite{Kainth98} The H-RPA
agreed better with the data for low temperatures ($T_{\rm e} \lesssim
T_{\rm F}/2$, the Fermi temperature) than did the RPA, suggesting that
exchange-correlation effects are important even for the comparatively
high 2DEG densities used; but for higher $T_{\rm e}$, the RPA gave a
better fit, suggesting that at higher temperatures ($T_{\rm e}
\gtrsim T_{\rm F}/2$) exchange-correlation effects become less
significant and the RPA the more appropriate approach.

The temperature dependence of the Coulomb drag between two 2DEGs has
also been employed recently to study the effects of electron
correlations in bilayer systems.\cite{Hill97,Noh98} The drag mechanism
was attributed to Landau damping of the
AP,\cite{Flensberg94,Flensberg95} but although inclusion of a 0~K
Hubbard local field correction provided a better description of the
data than that given by the RPA,\cite{Hill97} overall agreement
between experiment and theory was not achieved.  In both these Raman
and Coulomb drag studies\cite{Kainth98,Hill97} it was suggested that
using a static local field correction which was $T_{\rm e}$-dependent
might be important, and the present work addresses this point in
detail by comparing experimental data with corresponding calculations.

We present here Raman scattering measurements of the Landau damping 
induced width of the AP as a function of $T_{\rm e}$, and hence obtain 
a measure of the overlap of the AP with the SPE continuum and of the 
importance of exchange and correlation in these systems.  The data are 
modelled using calculations (taking full account of the finite 
thickness of the quasi-2DEGs, and of phenomenological damping) in: (i) 
the RPA, which is expected to provide a good description when exchange 
and correlation are not important; (ii) the RPA with a 0 K Hubbard 
local field correction to account for exchange-correlation effects 
(H-RPA); (iii) the self-consistent STLS approach, {\it modified to 
include non-zero $T_{\rm e}$}.  Most previous theoretical work on 
exchange-correlation effects in electron bilayers has assumed $T_{\rm 
e}=0$~K ({\textit e.g.} Refs.  \onlinecite{Gold92} to 
\onlinecite{Liu96}), but a key feature of our experimental work is 
that we have raised $T_{\rm e}$ substantially ($\sim T_{\rm F}$), and 
it is important to account for these observations.  The capacity of 
the STLS approach to include exchange-correlation effects more 
reliably that the H-RPA at low $T_{\rm e}$ is not especially relevant 
here; indeed, at the relatively high 2DEG densities and momentum range 
considered here, the RPA and the STLS approach should produce similar 
results.\cite{STLS} What is important is that the STLS can be readily 
modified to account for the effects of non-zero $T_{\rm e}$ on the 
static local field corrections.

As will be shown, there remain discrepancies between our experimental
results and those predicted by these theoretical descriptions of
electron fluids at non-zero $T_{\rm e}$, and this is one of the
central conclusions of this work.  In particular, the static STLS and
more generally {\em{all}} static local field factors
fail to describe effectively the effects of exchange and correlation
in a non-degenerate 2DEG. However, our results do suggest that a
{\it dynamic} local field correction could provide an effective description.

The theoretical approaches employed are considered in
Section \ref{Theorysec} and Appendix \ref{STLSapp}, in which we also
provide an overview of the RPA, H-RPA and STLS approximations for
electron bilayers.  The effects of non-zero temperature and 2DEG
thickness on the STLS local field factors and pair-correlation
functions are considered in Section \ref{Theoryres}.  The calculated
forms of the lineshapes for Raman scattering from the bilayer plasmons
at non-zero temperatures are discussed in Section \ref{Ramanth}.
Details of the experimental measurements are provided in Section
\ref{expt} and in Section \ref{Res} we present the results of our
Raman scattering measurements and discuss them in terms of the
theoretical models.

\section{Theory}
\label{Theorysec}

\subsection{The RPA for an electron bilayer}

Our starting point for a theoretical description of the excitation
spectrum of a double quantum well system is the RPA, which has been
used extensively for the study of the collective excitations in
low-dimensional electron systems,\cite{Stern67,Fasol89,Richards90a}
including electron
bilayers.\cite{DasSarma81,Vitlina81,Jain87,Santoro88} Throughout, we
will be concerned with excitations of angular frequency $\omega$ and
wavevector (parallel to the planes of the 2DEGs) ${\bf q}$.  We assume
that the two 2DEGs, which will be labelled $i$ and $j$, are isotropic.

The charge-density fluctuation $\delta \rho_{i}(q,\omega)$ in layer
$i$ induced by an external potential ${\phi_{j}}^{\rm ext}(q,\omega)$
in layer $j$ is given by
\begin{equation}
\delta \rho_{i}(q,\omega) = \chi_{ij}(q,\omega) {\phi_{j}}^{\rm
ext}(q,\omega)\ \ ,
\label{chi_define}
\end{equation}
\noindent where $\chi_{ij}(q,\omega)$ are the elements of the bilayer
density response function $\overline{\overline{\chi}}(q,\omega)$.
Within the RPA (and the STLS approximation), this is given by (see
Ref.  \onlinecite{Zheng94} for further details)
\begin{equation}
\overline{\overline{\chi}} = \dfrac{1}{1 - \widetilde{v}_{12}
\widetilde{v}_{21} \chi_{1} \chi_{2}} \left(\begin{array}{cc} \chi_{1}
& \widetilde{v}_{12} \chi_{1} \chi_{2} \\
\widetilde{v}_{21} \chi_{1} \chi_{2} & \chi_{2}
\end{array}
\right) \ \ ,
\label{eq:def_chi(ij)}
\end{equation}
\noindent where $\widetilde{v}_{ij}(q)$ is an effective interaction
between an electron in layer $i$ and an electron in layer $j$; within
the RPA this is the Coulomb interaction $v_{ij}(q)$.
$\chi_{i}(q,\omega)$ is the density response function for a single,
isolated layer:
\begin{equation}
\chi_{i}=\frac{\chi_{i}^{0}}{1 - \widetilde{v}_{ii}\chi_{i}^{0}}\ \ ,
\label{eq:def_chi(i)}
\end{equation}
\noindent where $\chi_{i}^{0}(q,\omega)$ is the 2D Lindhard
non-interacting electron polarisability for layer $i$.
$\chi^{0}(q,\omega)$ for a parabolic conduction band is given
by\cite{Stern67}
\begin{equation}
\chi^{0}(q,\omega;\mu,T) = \frac{2m}{\hbar^{2}} \int{\frac{{\rm d}{\bf
k}}{(2\pi)^{2}}} \frac{f^{0}({\bf k}+{\bf q}/2) - f^{0}({\bf k}-{\bf
q}/2)} {{\bf k}\cdot{\bf q}-m(\omega + i\gamma)/\hbar}\ \ ,
\label{Lindhard}
\end{equation}
\noindent where $f^{0}({\bf k})$ is the Fermi-Dirac function for
(temperature dependent) chemical potential $\mu$ and $\gamma$ is a
phenomenolgical damping parameter which we take to be the inverse of
the single particle relaxation time ($\gamma=1/2\tau_{s}$).  Mermin
has shown that for non-zero $\gamma$ the above form for
$\chi^{0}(q,\omega)$ does not strictly conserve particle
number;\cite{Mermin70} however, the sample quality in the present work
was such that $\gamma$ was sufficiently small to ensure that the above
approximation (Eq.  \ref{Lindhard}) is adequate.

To evaluate $\chi^{0}(q,\omega)$ for non-zero temperatures we follow
the method set out by Maldague\cite{Maldague78} and rewrite the
integral (Eq.  \ref{Lindhard}) as:\cite{Flensberg95}
\begin{equation}
\chi^{0}(q,\omega;\mu,T) =
\int_{0}^{\infty}\chi^{0}(q,\omega;\mu^{\prime},T=0) \dfrac{1}{4
k_{\rm B}T \cosh^{2} [(\mu-\mu^{\prime})/(2k_{\rm B}T)]} ~{\rm
d}\mu^{\prime}\ \ .
\label{Maldague}
\end{equation}
\noindent The polarisability at zero temperature with non-zero damping
is given by\cite{OConnell88}
\begin{equation}\label{A}
\chi^{0}(q,\omega;\mu=E_{\rm F},T=0) = \frac{2m}{h^{2}}~\frac{k_{\rm
F}}{q} \left[B\left(-\frac{q}{2k_{\rm F}} - \frac{\omega}{v_{\rm F}q}
- \frac{i \gamma}{v_{\rm F}q}\right) - B\left(\frac{q}{2k_{\rm F}} -
\frac{\omega}{v_{\rm F}q} - \frac{i \gamma}{v_{\rm F}q}\right)\right]\
\ ,
\label{chi_explicit}
\end{equation}
\noindent where
\begin{eqnarray}\label{B}
\nonumber B(x - i \alpha) = \frac{1}{\sqrt{2}} \left[\sqrt{2}x -
{\rm{sgn}}(x)\left((x^{2}-\alpha^{2}-1) +
\sqrt{(x^{2}-\alpha^{2}-1)^{2} +
4{x}^{2}\alpha^{2}}\right)^{1/2}\right] \\
+ \frac{i}{\sqrt{2}}\left[-\sqrt{2}\alpha + \left((1+\alpha^{2}-{x}^2)
+ \sqrt{(1+\alpha^{2}-{x}^2)^{2} +
4{x}^{2}\alpha^{2}}\right)^{1/2}\right]\ \ .
\end{eqnarray}
\noindent Using Eqs.  \ref{A} and \ref{B}, Eq.  \ref{Maldague} can be
evaluated numerically to gives a form for the non-interacting electron
polarisability at non-zero temperatures with damping.

\begin{figure}
\begin{center}
\epsfxsize = 12 cm 
\epsfbox{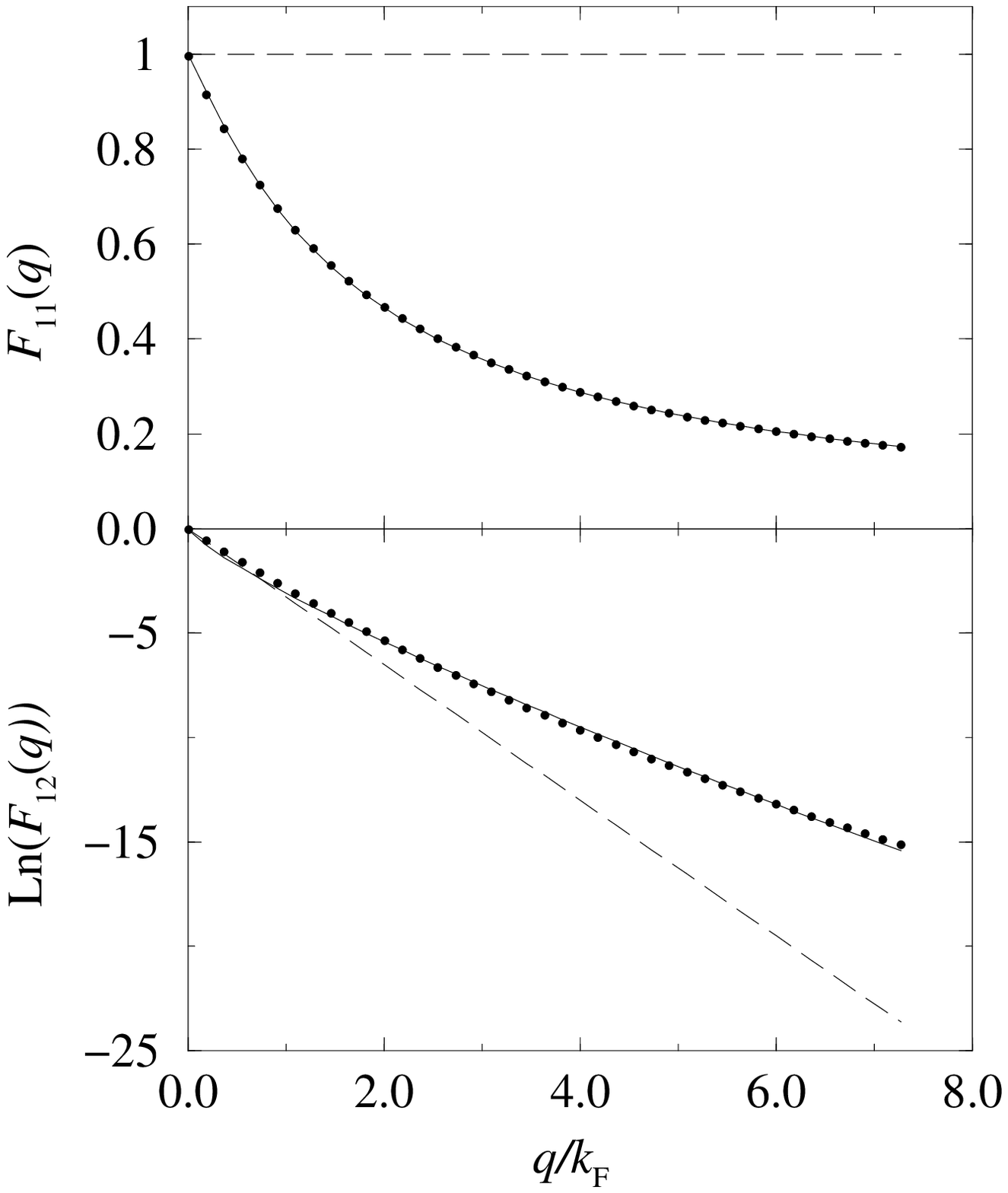}
\end{center}
\caption{The intra-layer and inter-layer Coulomb form factors $F_{11}$
and $F_{12}$, determined with quantum size effects taken into account
($\bullet$).  Fits using the functional forms given by Eq.  \ref{Fij}
are indicated by the solid lines.  The dashed lines give the form
factors $F_{11}^{0}(q)=1$ and $F_{12}^{0}(q)=\exp(-qd)$ for the
strictly two-dimensional approximation.  Note the logarithmic scale
for $F_{12}$.}
\label{Formfig}
\end{figure}

For a realistic system, it is also necessary to account for the finite
spatial extent of the electron envelope wavefunctions along the
confinement direction for the 2DEGs.  This is achieved with the
Coulomb form factors $F_{ij}$, defined by $F_{ij}(q)=v_{ij}(q)/v(q)$,
where $v(q)=e^{2}/2\epsilon\epsilon_{0}q$ is the 2D Coulomb
interaction and $\epsilon$ is the effective dielectric function for
the media surrounding the bilayer.  Interactions between plasmons and
phonons were included using a frequency dependent dielectric
constant.\cite{Kainth99} Using envelope wavefunctions $\psi_{i}(z)$,
determined (for sample structural parameters corresponding to the
experimental sample discussed in Section \ref{expt}) from a
self-consistent solution of the Poisson and Schr\"{o}dinger equations
(see Ref. \onlinecite{Kainth99} for further details), the form
factors
\begin{equation}
F_{ij}(q) = \frac{v_{ij}(q)}{v(q)} = \int{\ {\rm d}z \int{\ {\rm
d}z'e^{-q|z-z'|}|\psi_{i}(z)|^{2}|\psi_{j}(z')|^{2}}}
\label{Form_factor}
\end{equation}
\noindent were evaluated numerically for wavevectors up to $q =
6k_{\rm F}$ and fitted to functions of the forms
\begin{eqnarray}
\nonumber F_{11}(q) &=& 1/(1 + b q^{m}) \\
F_{12}(q) &=& \exp(-d ^{*} q^{n}) \ \ ,
\label{Fij}
\end{eqnarray}
\noindent where $b$, $d^{*}$, $m$ and $n$ are fitting parameters.  We
show in Fig.  \ref{Formfig} the $q$ dependence of $F_{11}$ and
$F_{12}$ determined from Eq.  \ref{Form_factor}, together with the
fits obtained using Eq.  \ref{Fij}.  The form factors for the strictly
two-dimensional approximation, $F_{11}^{0}(q)=1$ and
$F_{11}^{0}(q)=\exp(-qd)$, are also shown for comparison ($d$ is the
effective separation of the 2DEGs\cite{Kainth99}); for large
wavevector $q$ the 2D approximations for the form factors fail
significantly.

The functional forms for the Coulomb form factors given in Eq.
\ref{Fij}, together with $\chi^{0}(q,\omega;\mu,T)$ from Eq.
\ref{Maldague}, provide $\chi_{i}$ (Eq.  \ref{eq:def_chi(i)}) and
$\overline{\overline{\chi}}(q,\omega;\mu,T)$(Eq.
\ref{eq:def_chi(ij)}) within the RPA.

\subsection{The optic and acoustic plasmons of an electron bilayer}

The energies of the collective charge-density excitations, the
plasmons, of the electron bilayer system are given (in the absence of
damping, $\gamma = 0$) by the poles of the the density response
function $\overline{\overline{\chi}}(q,\omega)$.  Direct insight into
the physical character of the charge-density excitation spectrum is
obtained by diagonalising $\overline{\overline{\chi}}(q,\omega)$, to
give, as the diagonal elements, the density response functions for
charge-density fluctuations in which the fluctuations in the two
layers are in-phase ($\chi_{+}$) and out-of-phase ($\chi_{-}$)
:\cite{Liu96}
\begin{equation}
\chi_{\pm} = \dfrac{2}{\left(\dfrac{1}{\chi_{1}} +
\frac{1}{\chi_{2}}\right) \pm \sqrt{\left(\dfrac{1}{\chi_{1}} -
\frac{1}{\chi_{2}}\right)^{2} + 4 \widetilde{v}_{12}^{2}}}\ \ .
\label{eq:chi_diag}
\end{equation}
\noindent The poles of $\chi_{+}(q,\omega)$ and $\chi_{-}(q,\omega)$
give the optic plasmon (OP) and acoustic plasmon (AP) respectively.

More experimentally realistic response functions, with non-zero
damping ($\gamma \neq 0$) allow us to calculate the cross-section
$R(q,\omega)$ for Raman scattering by charge density fluctuations from
the imaginary part of the density response functions\cite{LSSIV}
\begin{equation}
R_{\pm}(q,\omega) \propto - (n(\omega)+1) 
{\rm Im}[\chi_{\pm}(q,\omega)]\ \ ,
\label{Raman}
\end{equation}
\noindent and hence the expected Raman lineshapes for the two plasmon 
modes; the energies of the plasmon modes are determined from the peaks 
in the intensities $R_{\pm}(q,\omega)$.  The factor $(n(\omega)+1)$ in 
Eq.  \ref{Raman}, where $n(\omega)$ is the Bose-Einstein occupation 
function, was not considered in Ref.  \onlinecite{Kainth98}, the 
result of which was to slightly underestimate the asymmetry and widths 
of the theoretically predicted Raman lineshapes.  In this work we will 
not include this factor in our theoretical analysis but rather correct 
experimental spectra for its contribution; in particular this allows 
for a more meaningful comparison between experiment and theory of peak 
widths, which give a measure of plasmon damping.

\subsection{The STLS approximation}
\label{STLSsec}
Although the RPA has been very successful in providing a description
of the excitation spectra of 2DEG systems, it takes no account of the
effects of exchange and correlation.  This failing of the RPA becomes
clear if we consider the instantaneous pair correlation functions
$g_{ij}(r)$, which give the probability of finding an electron at an
in-plane position ${\bf r}$ in layer $j$, given that there is an
electron at the origin in layer $i$.\cite{Szymanski94} This is given
by:\cite{Zheng94,Szymanski94}
\begin{equation}
g_{ij}(r)=1+\frac{1}{\sqrt{N_{i }N_{j}}}\int{e^{i {\bf q} \cdot {\bf
r}}[S_{ij}(q) - \delta_{ij}]}\, \frac{{\rm d}{\bf q}}{(2\pi)^{2}}\ \ ,
\label{eq:def_g(ij)}
\end{equation}
\noindent where $N_{i}$ is the equilibrium number density in layer
$i$.  The static structure factors $S_{ij}(q)$ are related to the
density response function $\overline{\overline{\chi}}(q,\omega)$ by
the fluctuation dissipation theorem:\cite{Tanaka86}
\begin{equation}
S_{ij}(q) = \frac{-\hbar}{2\pi \sqrt{N_{i}N_{j}}}\ {\cal P}
\int\limits_{-\infty}^{\infty} {\coth \left( \frac{\hbar \omega}
{2k_{\rm B} T}\right)~{\rm Im}[\chi_{ij}(q,\omega)]~{\rm d}\omega} \ \
,
\label{eq:def_S(ij)}
\end{equation}
\noindent where $\cal{P}$ denotes the principal value of the integral.

The RPA gives the unphysical result of negative $g_{ij}(r)$ for small
$r$, implying a much deeper exchange correlation hole than is
physically realistic and hence overestimating significantly the
correlation energy, the energy of interaction between the electron and
its associated correlation hole.\cite{Zheng94} To correct for this,
Singwi \textit{et al.} allowed for a local depletion of the electron
density around any given particle.\cite{Singwi68,STLS} For the two
layer case this ansatz leads to an effective interaction between the
responding electron in layer $i$ and the induced charge in layer $j$
of the form: %
\begin{equation}
\widetilde{v}_{ij}(q) = v_{ij}(q) [1 - G_{ij}(q)] \ \ ,
\label{eq:linear response3}
\end{equation}
\noindent where the local field factors $G_{ij}(q)$ are given by
\begin{equation}
G_{ij}(q) = \frac{-1}{\sqrt{N_{i}N_{j}}} \int{\frac{\,{\rm d}{\bf
k}}{2\pi^{2}}\frac{{\bf q} \cdot {\bf k}}{q^{2}}
\frac{v_{ij}(k)}{v_{ij}(q)}\,[S_{ij}(|{\bf q} - {\bf
k}|)-\delta_{ij}]}\ \ .
\label{eq:def_G(ij)}
\end{equation}
\noindent For an electron bilayer system, the $G_{ij}({\bf q})$ are
generally positive and so the inclusion of exchange-correlation
creates a `hole' in the induced charge, reducing the strength of the
interaction between the responding electron and the induced charge.

From the local field factors $G_{ij}(q)$ we may determine
$\overline{\overline{\chi}}(q,\omega)$, the density response function
of the electron bilayer system, using the modified effective
potentials $\widetilde{v}_{ij}$ (Eq.  \ref{eq:linear response3}) in
Eqs.  \ref{eq:def_chi(ij)}, \ref{eq:def_chi(i)} and \ref{eq:chi_diag},
which in turn allows a determination of the structure factors
$S_{ij}(q)$ through the fluctuation dissipation theorem (Eq.
\ref{eq:def_S(ij)}).  Thus the determination of the local field
factors within the STLS approximation involves the self-consistent
evaluation of $\chi_{ij}(q,\omega)$, $S_{ij}(q)$ and $G_{ij}(q)$ (Eqs.
\ref{eq:def_chi(ij)}, \ref{eq:def_S(ij)} and \ref{eq:def_G(ij)}).

Our method for the self-consistent evaluation of these STLS local
field factors $G_{ij}$ for a two layer system at finite temperature is
described in detail in Appendix \ref{STLSapp}, and mirrors the
approaches of Tanaka and Ichimaru, who studied a one component plasma
system at finite temperature in three dimensions within the STLS,
principally with a view to modelling stellar cores,\cite{Tanaka86} and
of Schweng and Bohm who extended this work to examine a one component
electron liquid in two dimensions.\cite{Schweng94} Note that for
simplicity we assume that the two layers are of equal density i.e.
$N_{i}=N_{j}=N$.  This approximation is justified for the sample
studied experimentally, where any asymmetries between the wells were
minimal; calculated Fermi energies for the two wells were within
$0.5\%$ and no experimental asymmetries were observed in,
\textit{e.g.} photoluminescence measurements or in the dispersive
behaviour of the acoustic and optic plasmon modes at low
temperature.\cite{Kainth98,Kainth99}

The local field factors $G_{ij}$ evaluated within the self-consistent
scheme outlined in Appendix \ref{STLSapp} are then used to determine
the diagonalised density response functions $\chi_{\pm}(q,\omega)$
(Eq.  \ref{eq:chi_diag}), determined with the Lindhard polarisability
given by Eqs.  \ref{Maldague} and \ref{chi_explicit}.  The RPA result
is recovered if the local-field factors $G_{ij}(q)$ are set to zero.
These density response functions allow us to determine (through Eq.
\ref{Raman}), as a function of electron temperature $T_{\rm e}$, the
Raman scattering lineshapes for the acoustic and optic plasmons of the
electron bilayer, which we then compare with our experimental results.

\subsection{The Hubbard approximation --- H-RPA}

The simplest way to go beyond the RPA is to avoid the self-consistent
scheme set out in the previous Section and to use instead the
Hartree-Fock approximation for the static intra-layer structure factor
$S_{ii}(q)$,\cite{Jonson76} which accounts only for the presence of
the Pauli hole around each electron at $T=0$.  The effects of non-zero
temperature and contributions to the local field from intra- and
inter-layer Coulomb correlations are therefore neglected in this
approximation --- the Hubbard correction.  The resultant integral for
$G_{ij}(q)$ (Eq.  \ref{eq:def_G(ij)}) can then be evaluated explicitly
(within an approximation) to yield:\cite{Jonson76}
\begin{equation}
G_{ij}^{\rm H}(q)=\left\{
        \begin{array}{lcc}
                \dfrac{q}{2\sqrt{q^{2}+k_{{\rm F}i}^{2}}} & $if$ & i= j\\
                \ 0 & $if$ & i\neq j \\
        \end{array}
    \right.
\label{eq:def_G(hubbard)}
\end{equation}
\noindent where $k_{{\rm F}i}$ is the Fermi wavevector in layer $i$.

This Hubbard local field correction $G_{ij}^{\rm H}(q)$ was used to
account for the effects of electron correlations in the analysis of
the Coulomb drag measurements of Ref.  \onlinecite{Hill97} and in our
preliminary report on the temperature dependent Landau damping of the
acoustic plasmon in an electron bilayer.\cite{Kainth98}

\section{Calculations of correlations in electron bilayers}
\label{Theoryres}
We present in this section the dependence of the STLS local field
corrections $G_{ij}(q)$ and pair correlation functions $g_{ij}(r)$ of
an electron bilayer on temperature and 2DEG thickness.  For all the
calculations presented here, we have taken the parameters for the
electron bilayer to be those for the sample investigated
experimentally (see Section \ref{expt}).  In fact the results show
that the effects of temperature and non-zero layer thickness on the
inter- and intra-layer STLS local field corrections and correlations
are not significant for this sample.

\subsection{Temperature dependence of local field factors}

Our calculations of the temperature dependent local field factors
$G_{ij}(q)$ for electron bilayers reproduce the results of earlier
calculations in the limit of zero temperature and layer thickness.  In
particular we see similar behaviour of the local field factors as
functions of number density $N$ (or equivalently $r_{\rm s} = (N \pi
a_{\rm B}^{2})^{-1/2}$) and interlayer separation ($d$), to that
described by Liu \textit{et al.}\cite{Liu96}

\begin{figure}
\begin{center}
\epsfxsize = 12 cm 
\epsfbox{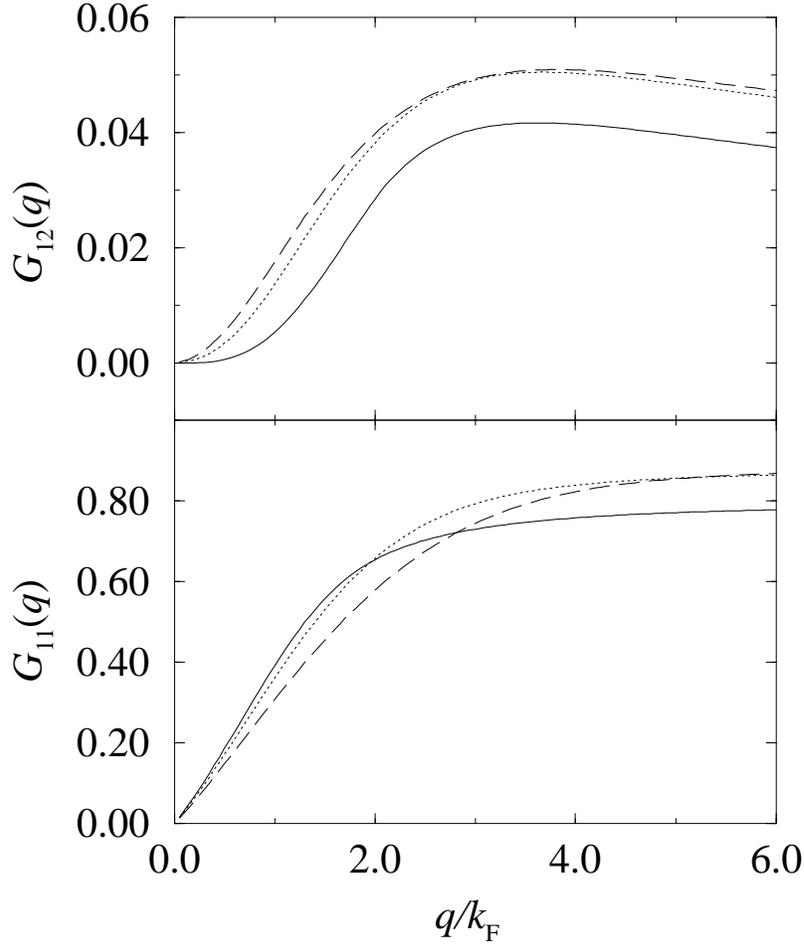}
\end{center}
\caption{Variation of (a) the intra-layer $G_{11}(q)$ and (b)
inter-layer $G_{12}(q)$ local field factors with wavevector $q$, for
$\theta = 0.05$ (solid line), 1.0 (dotted line) and 2.0 (dashed line);
$\theta = T_{\rm e}/T_{\rm F}$, $T_{\rm F} = 78$~K.}
\label{Loc_Fld_Temp}
\end{figure}

\begin{figure}
\begin{center}
\epsfxsize = 12 cm 
\epsfbox{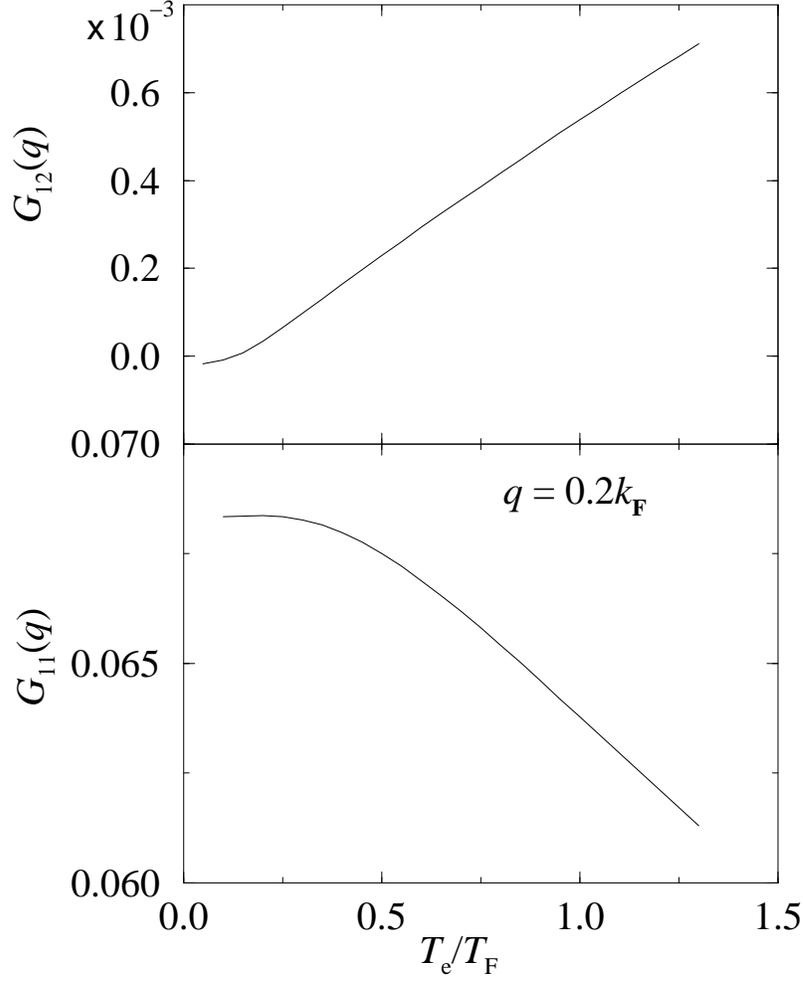}
\end{center}
\caption{Variation of the intra-layer and inter-layer local field
factors $G_{11}(q)$ and $G_{12}(q)$ with electron temperature $T_{\rm
e}$, for $q = 0.2~k_{\rm F}$.  Note the scaling factor of $10^{3}$ for
$G_{12}$.}
\label{GvsT}
\end{figure}

In Fig.  \ref{Loc_Fld_Temp} we show the $q$ dependence of the local
field corrections $G_{11}$ and $G_{12}$ as a function of temperature
($\theta = T_{\rm e}/T_{\rm F}$).  Although the wavevector range (to
$6~k_{\rm F}$) is not relevant to our experimental work, it was
necessary to determine $G_{ij}$ for large $q$ in the self-consistent
STLS scheme (Eq.  \ref{eq:def_G(ij)}), and we present these results
for completeness and to illustrate the integrity of our calculations.
The dependence on temperature of $G_{11}$ and $G_{12}$ for $q =
0.2~k_{\rm F}$ (the typical magnitude of wavevector accessible in the
Raman experiments) is illustrated in Fig.  \ref{GvsT}.  Although
$G_{12}$ increases significantly with temperature, the magnitude of
this inter-layer local field factor is negligible for this wavevector.
$G_{11}$, on the other hand, is much larger and is reduced (although
by only $\sim 5 \%$) on increasing the temperature from absolute zero
to the Fermi temperature.

\begin{figure}
\begin{center}
\epsfxsize = 12 cm 
\epsfbox{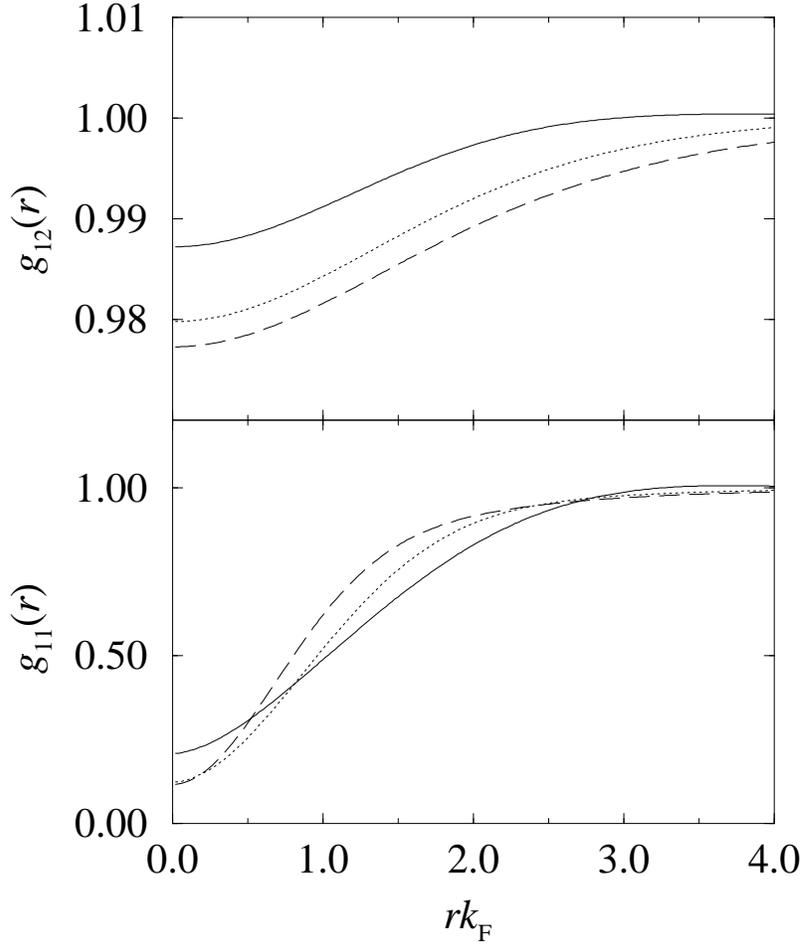}
\end{center}
\caption{Variation of the pair correlation functions $g_{11}(r)$ and
$g_{12}(r)$, with in-plane separation $r$, for $\theta = 0.05$ (solid
line), 1.0 (dotted line) and 2.0 (dashed line); $\theta = T_{\rm
e}/T_{\rm F}$, $T_{\rm F} = 78$~K.}
\label{Pcf_Temp}
\end{figure}

Fig.  \ref{Pcf_Temp} shows the dependences of $g_{11}(r)$ and
$g_{12}(r)$, the pair correlation functions, on $r$, for several
temperatures; the minimum in the pair correlation functions describes
the exchange-correlation hole.  The dependence on temperature of the
intra-layer pair correlation function $g_{11}(r)$ is similar to that
for a single layer;\cite{Schweng94} as $T_{\rm e}$ increases the
radius of the exchange-correlation hole is reduced, accompanied by a
slight deepening of the hole, in accordance with screening sum
rules:\cite{Szymanski94}
\begin{equation}
\int{[1 - g_{ij}(r)]{\rm d}{\bf r}} = \delta_{ij}\ \ .
\label{screeningsum}
\end{equation}
\noindent The inter-layer pair correlation function $g_{12}(r)$ for
these comparatively high densities ($r_{\rm s}\sim1.3$) is, as
expected, close to 1.

\subsection{Quantum size effects}

\begin{figure}
\begin{center}
\epsfxsize = 12 cm 
\epsfbox{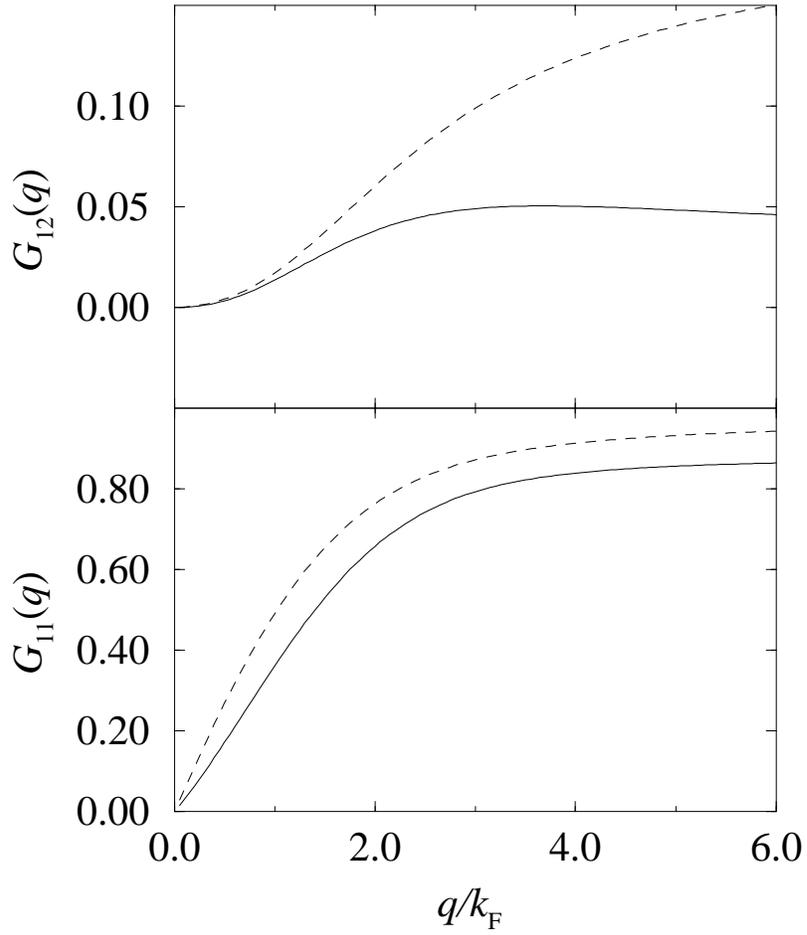}
\end{center}
\caption{A comparison of the $T = 0$ local field corrections
$G_{11}(q)$ and $G_{12}(q)$ determined in the ideal 2D approximation
(dashed line) and with quantum size effects of the quasi-2DEGs taken
into account (solid line).}
\label{Layer_LFC}
\end{figure}

In Fig.  \ref{Layer_LFC} we show the $q$ dependence of the intra- and
inter-layer local field corrections $G_{11}$ and $G_{12}$ determined
with the finite thickness effects of the quantum wells taken into
account (using the functional forms for $F_{11}$ and $F_{12}$ given in
Eq.  \ref{Fij}), which we compare with the corresponding results
determined within the strictly two-dimensional approximation (for
which $F_{11}^{0}(q) = 1$ and $F_{12}^{0}(q) = \exp(-qd)$).  Our
results are in line with those for one layer systems,\cite{Jonson76}
in that correlations are reduced in going from the ideal to the
quasi-2D regime.  The greatest discrepancies occur for large
wavevector $q$, where the 2D approximations for the Coulomb potentials
fail significantly (see Fig.  \ref{Formfig}).  In particular, the
inter-layer correlations for a real system do not continue to increase
as a function of $q$, but instead reach a maximum and then decline.

\begin{figure}
\begin{center}
\epsfxsize = 12 cm 
\epsfbox{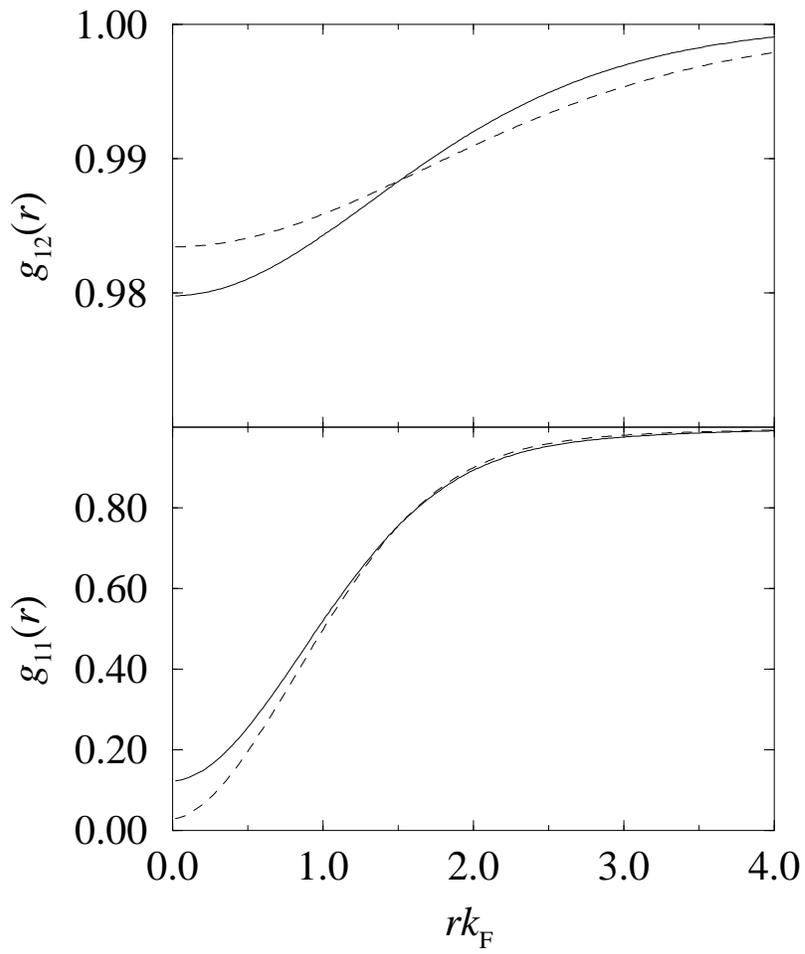}
\end{center}
\caption{The pair correlation functions, $g_{11}(r)$ and $g_{12}(r)$,
as for Fig.  \ref{Layer_LFC}.}
\label{Layer_PCF}
\end{figure}

Fig.  \ref{Layer_PCF} shows a similar comparison between the quasi-2D
and ideal-2D cases for the intra- and inter-layer pair correlation
functions, $g_{11}$ and $g_{12}$.  Compared with the ideal-2D case,
electron-electron effects are expected to be reduced in the quasi-2D
regime because the spread of charge in the $z$-direction reduces the
intra-layer Coulomb interactions; so there should be a shallower
intra-layer exchange correlation hole in the quasi-2D regime, as
confirmed in the calculations.  This shallowing of the intra-layer
hole is accompanied by a deepening of the inter-layer hole, as
electrons in different layers do not in general want to sit opposite
each other.  These effects of the charge spreading in the
$z$-direction are significant for this density ($r_{\rm s} \sim 1.3$)
because the mean distance between the electrons ($\sim 1/k_{\rm F}
\sim 90~{\rm \AA}$) is significantly less than the layer width.

\section{Raman scattering from charge-density fluctuations}
\label{Ramanth}

\begin{figure}
\begin{center}
\epsfxsize = 12 cm 
\epsfbox{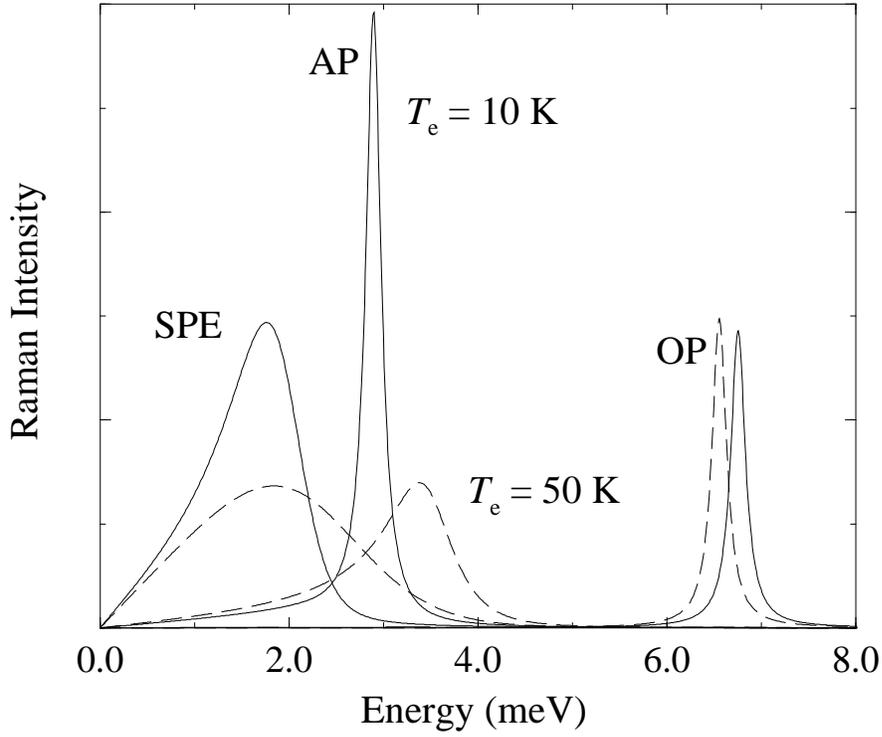}
\end{center}
\caption{Theoretical Raman spectra for charge-density fluctuations 
(solid lines) and SPE (dashed lines) for $T_{\rm e} = 10~{\rm K}$ 
($0.13\, T_{\rm F}$) and $T_{\rm e} = 50~{\rm K}$ ($0.64\, T_{\rm 
F}$).  The damping parameter $\gamma = 0.09~{\rm meV}$ and the 
wavevector $q = 1.6\times 10^{5}~{\rm cm}^{-1}$.  The occupation 
factor $(n(\omega)+1)$ in Eq.  \ref{Raman} has been omitted. Peaks due 
to the acoustic (AP) and optic (OP) plasmons are present in the 
charge-density fluctuation spectra.  With increasing $T_{\rm e}$, the 
SPE continuum spreads to higher energies, overlapping the AP energy 
and leading to an asymmetric broadening of the AP peak, resulting from 
Landau damping.}
\label{ThLDspectra}
\end{figure}

Fig.  \ref{ThLDspectra} shows calculated spectra for Raman scattering 
by the AP (derived from $-{\rm Im}[\chi_{-}(q,\omega)]$), the OP (from 
$-{\rm Im}[\chi_{+}(q,\omega)]$) and the SPE (from $-{\rm 
Im}[\chi^{0}(q,\omega)]$) for $T_{\rm e}=10$ K and 50 K. As $T_{\rm 
e}$ increases, the SPE spectrum spreads upwards in energy, as 
expected.  This results in an increasing population of electrons 
travelling with the same velocity as the phase velocity of the AP. 
These electrons are able to exchange energy efficiently with the 
plasmon and so enhanced scattering of the AP by SPE occurs, resulting 
in an asymmetric broadening of the AP peak on the low-energy side, a 
clear signature of Landau damping.\cite{RichardsInAs} The degree of AP 
broadening, and its asymmetry, can be characterised (this will be 
discussed later) in terms of the half-widths on the higher and lower 
energy sides of the peak.  In these calculations the occupation factor 
$(n(\omega)+1)$ in the Raman intensity (Eq.  \ref{Raman}) has been 
omitted, to highlight the effects of Landau damping. The effect of 
this factor, which becomes more significant with increasing 
temperature, is to increase the Raman intensity at low Raman shifts, 
leading to a further slight increase in asymmetry of the AP peak.

Note that the OP peak, at rather higher energy, does not overlap the 
SPE continuum, and is not detectably broadened by Landau damping, even 
at 50 K. Fig.  \ref{ThLDspectra} also shows the upward drift in the 
energies of the AP and the OP with increasing temperature, a well 
known classical effect to which we will return.

\section{Experimental measurements}
\label{expt}

\subsection{Sample details}

The sample investigated (T229; sample C in Refs.
\onlinecite{Kainth98} and \onlinecite{Kainth99}) consists of two GaAs
quantum wells of width $L_{\rm w} = 180~{\rm \AA}$, separated by an
Al$_{0.67}$Ga$_{0.33}$As barrier of width $L_{\rm b} = 125~{\rm \AA}$,
giving an effective inter-2DEG separation $d \sim 305~{\rm \AA}$.
These parameters are close to optimal for this study, in that $L_{\rm
b}$ is large enough to preclude inter-well tunnelling\cite{Kainth99}
(we have throughout assumed a quantum mechanically decoupled system),
while the inter-layer separation $d$ is small enough to lead to
significant inter-layer Coulomb interactions, which depress the energy
of the AP mode towards the SPE continuum and increase its
susceptibility to many-body effects.

The number densities $N_1$ and $N_{2}$ in the two wells under laser
illumination were obtained from a fit, calculated within the STLS
approximation, of the low-temperature experimentally determined AP and
OP dispersions.\cite{Kainth99} It was found that $N_1$ and $N_{2}$
were equal to within $\sim 5\%$, as expected since the two wells were
grown to be equivalent.  $N_{1}=N_{2}=(1.95 \pm 0.10) \times
10^{11}{\rm{cm}}^{-2}$; see Ref.  \onlinecite{Kainth99} for further
details.

The sample was mounted in a He-cooled cryostat and its temperature
controlled by electrical heating.

\subsection{Determination of the electron temperature}\label{Te}

\begin{figure}
\begin{center}
\epsfxsize = 10 cm 
\epsfbox{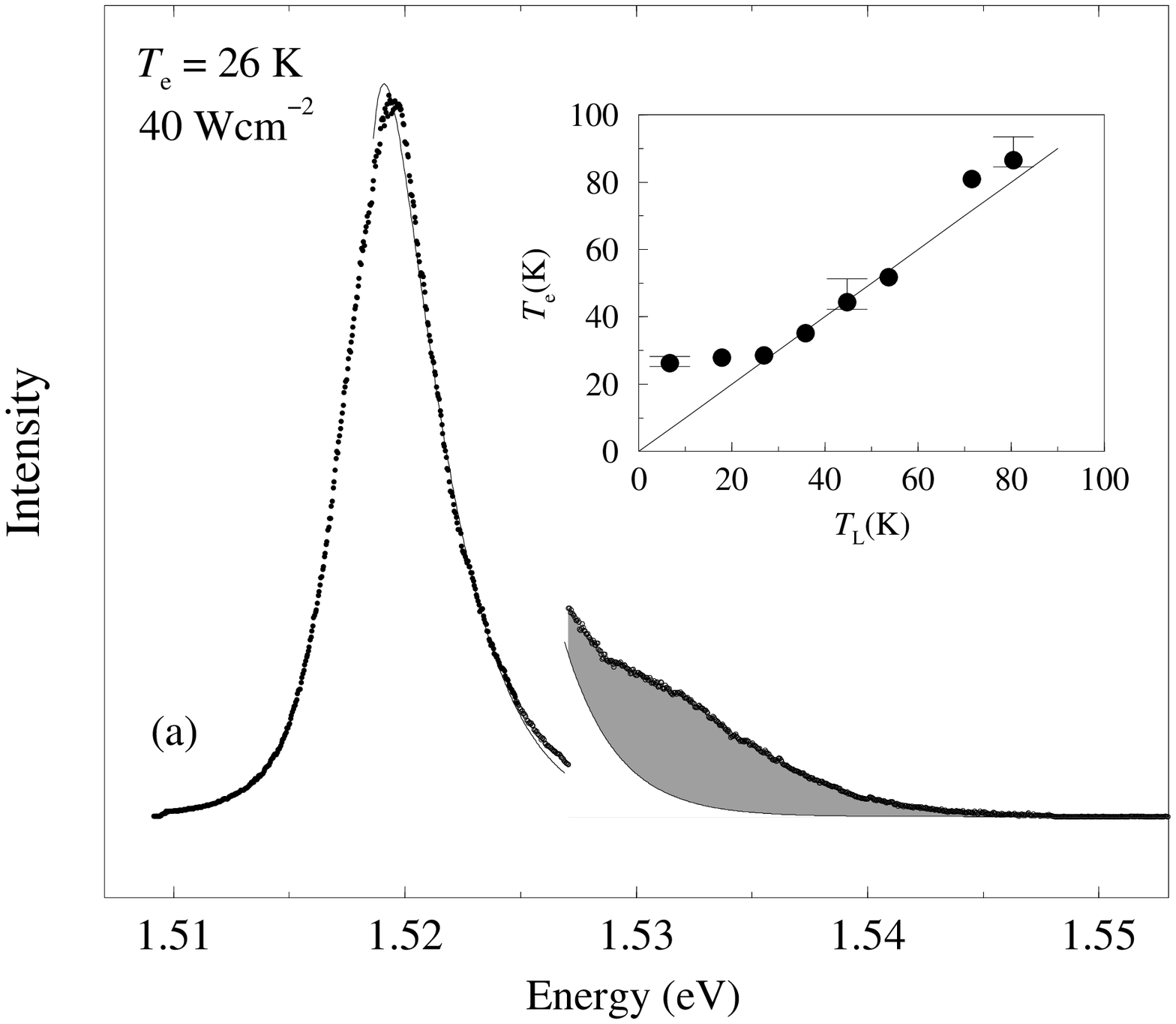}
\end{center}
\begin{center}
\epsfxsize = 10 cm 
\epsfbox{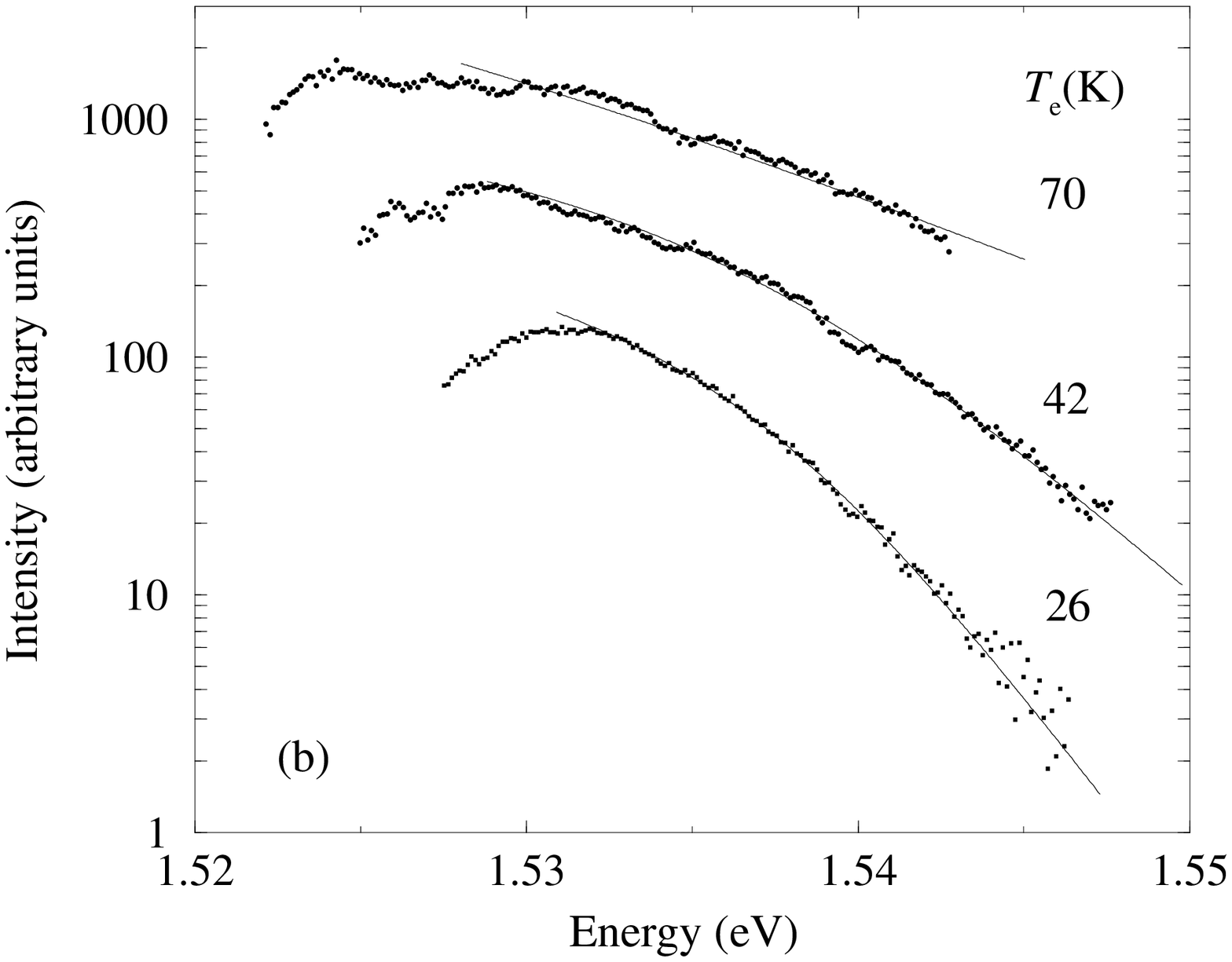}
\end{center}
\caption{(a) Bulk PL signal for $T_{\rm e}=26~{\rm K}$ and magnified
($\times 4)$ view of the quantum well PL. The solid line indicates the
fit described in the text.  The inset shows the variation of the 2DEG
temperature $T_{\rm e}$ with the lattice temperature $T_{\rm L}$.  (b)
The log of the quantum well PL intensity ($\bullet$) for various
temperatures $T_{\rm e}$ (spectra are displaced vertically for
clarity).  Fits to the data, using Eq.  \ref{QWPLtheory} are indicated
by the solid lines.}
\label{PLfig}
\end{figure}

The effect of the 2DEG temperature $T_{\rm e}$ on the plasmons was
studied by varying $T_{\rm L}$, the lattice temperature.  An accurate
determination of $T_{\rm e}$ was critical for the interpretation of
our experiment, and was achieved by carefully studying the quantum
well and bulk photoluminesence (PL) lineshapes, determined under the
same conditions as the Raman scattering measurements presented below;
i.e. for excitation photon energies $\geq 1.65$ eV and high powers (40
W cm$^{-2}$), with $T_{\rm e} \geq 25$ K. The strength of the bulk PL
signal from the GaAs buffer layer was found to be some 25 times
greater than the quantum well signal, as shown in Fig.
\ref{PLfig}(a); the lower energy signal is due to the bulk GaAs, and
the weak shoulder at $\sim 1.53$ eV is the quantum well PL (this
assignment has been confirmed by PL and PLE measurements performed
with lower power densities and electron temperatures).  Consequently
the PL tail from the bulk GaAs must be fitted and subtracted from the
experimental spectra to give the quantum well PL; a fit to the bulk PL
lineshape is shown in Fig \ref{PLfig}(a).

Having stripped the PL spectrum of the bulk PL, the remaining quantum
well PL lineshape ${\cal L}_{\rm QW}(\omega)$ was taken to be:
\begin{equation}
{\cal L}_{\rm QW}(\omega,T_{\rm e}) \propto \sum_{{\bf k}} f_{\rm
e}(E_{{\rm e},{\bf k}},T_{\rm e}) f_{\rm h}(E_{{\rm h},{\bf k}},T_{\rm
e}) \delta(E_{\rm g}^{\prime} + E_{{\rm e},{\bf k}} + E_{{\rm h},{\bf
k}} - \hbar\omega)\ \ ,
\label{QWPLtheory}
\end{equation}
\noindent where $f_{\rm e(h)}$ are the Fermi-Dirac occupation factors
for electrons (holes) with kinetic energy $E_{{\rm e},{\bf k}}$
($E_{{\rm h},{\bf k}}$) and wavevector ${\bf k}$.  $E_{\rm
g}^{\prime}$ is the quantum well band-gap and $\hbar\omega$ is the PL
photon energy.  Given the temperature range over which the
measurements were made, simplifications to the form of ${\cal{L}}_{\rm
QW}(\omega,T_{\rm e})$ (\textit{e.g.} by making assumptions about the
degeneracies of the electrons and holes) were not possible, and the
experimental quantum well PL lineshapes were fitted to ${\cal{L}}_{\rm
QW}(\omega)$ assuming only that: (i) Coulomb scattering renders the
electron and hole temperatures equal; (ii) the total extrinsic
electron density ($1.95 \times 10^{11}{\rm{cm}}^{-2}$ for each quantum
well) did not change with temperature; (iii) the photocreated and
thermal electron and hole population densities (used as fitting
parameters) were equal.  The electron and hole effective masses were
taken to be 0.07 and 0.18 respectively.  We show in Fig.
\ref{PLfig}(b) a logarithmic plot of the experimental QW PL for three
different temperatures, along with fits of the form given in Eq.
\ref{QWPLtheory}.  The data does not fit a straight line (the
exponential form often used) except well into the higher energy part
of the PL tails where the errors in the data are significant.

By fitting the bulk and QW PL lineshapes in this way, we were able to
determine the temperature $T_{\rm e}$ of the 2DEG for each Raman
measurement with a maximium error (for the highest temperatures) of
$\sim 5\%$.  We were able to derive a value for the lattice
temperature $T_{\rm L}$ from the form of the variation of $E_{\rm g}$
(determined from the bulk PL) with $T_{\rm L}$: \cite{EgvsT}
\begin{equation}
E_{\rm g} = E_{\rm g}(T_{\rm L} = 0) - \frac{(5.408\times 10^{-4} {\rm
eV})(T_{\rm L}/{\rm K})^{2}}{(T_{\rm L}/{\rm K})+204}
\label{bulk3}
\end{equation}
\noindent We show in the inset of Fig.  \ref{PLfig}(a) the variation
of $T_{\rm e}$ with the lattice temperature $T_{\rm L}$; the laser
illumination is seen to heat the electron gas substantially above the
lattice temperature for low $T_{\rm L}$ (due to the relatively high
photon energy and power densities used), with $T_{\rm e}$ and $T_{\rm
L}$ converging for $T_{\rm L} \gtrsim 30$~K.

\subsection{Raman scattering measurements}

Angle resolved Raman scattering measurements were made under incoming
resonance conditions, and with the polarisations of the incident and
scattered light parallel to allow the observation of
plasmons.\cite{LSSIV} The band-gap decreased with increasing lattice
temperature $T_{\rm L}$, so to maintain the conditions for an incoming
resonance the laser wavelength was increased correspondingly.  For the
low-temperature measurements the energy of the resonance used was
1.656 eV. Rather than determine a resonance profile for each
measurement temperature, the shift with increasing $T_{\rm e}$ in the
resonance energy from this value was determined using $T_{\rm L}$ and
the known variation of the semiconductor band-gap with temperature
(Eq.  \ref{bulk3}).

As the temperature increases the tail from the bulk bandgap PL
increases considerably, and can overwhelm the comparatively weak Raman
signals.  Fortunately the resonance utilised here was sufficiently far
from the bandgap PL to allow Raman measurements to be made over a
reasonably large temperature range.  The band-gap PL, and the hot PL
originating from the inter-band transition responsible for the Raman
resonance, limited the Raman measurements to $T_{\rm e} \lesssim 110$
K.

\subsection{Single particle lifetime}
\label{SdHsec}

In our calculation of the Raman scattering lineshapes for the plasmon
modes, the phenomenological damping constant $\gamma$ (see Eq.
\ref{Lindhard}) is taken to be the inverse of the 2DEG single particle
relaxation time $\tau_{\rm s}$, which we determined experimentally
from the dependence of the magnitude of Shubnikov-de Haas oscillations
with magnetic field; only one Shubnikov-de Haas oscillation frequency
was observed, indicating that the two quantum well densities were the
same.  The amplitude $(\Delta R)$ of Shubnikov-de Haas oscillations is
given by:\cite{Coleridge91}
\begin{equation}
\Delta R = 4 R_{0}X(T)\exp(-~\dfrac{\pi}{\omega_{\rm c}\tau_{\rm s}})\
\ ,
\label{LifshitzKoseivich}
\end{equation}
\noindent where $R_{0}$ is the zero field resistance, $\omega_{\rm c}$
the cyclotron frequency, and $X(T)$ a thermal damping factor given by:
\begin{equation}
X(T)=\dfrac{2\pi^{2}k_{\rm B}T/\hbar\omega_{\rm c}}
{\sinh(2\pi^{2}k_{\rm B}T/\hbar\omega_{\rm c})}\ \ .
\label{thermalfactor}
\end{equation}
\noindent For our samples, $\tau_{\rm s}$ was found to be of the order
of 3.5 ps for Shubnikov-de Haas oscillations measured at 4 K. Using
this value, we were able to fix the low temperature half width of the
plasmons as $\hbar \gamma = \hbar/2\tau_{\rm s} = 0.09~{\rm meV}$.

\section{Results}
\label{Res}

\subsection{Temperature dependent Raman scattering}
\label{Res1}

\begin{figure}
\begin{center}
\epsfxsize = 10 cm 
\epsfbox{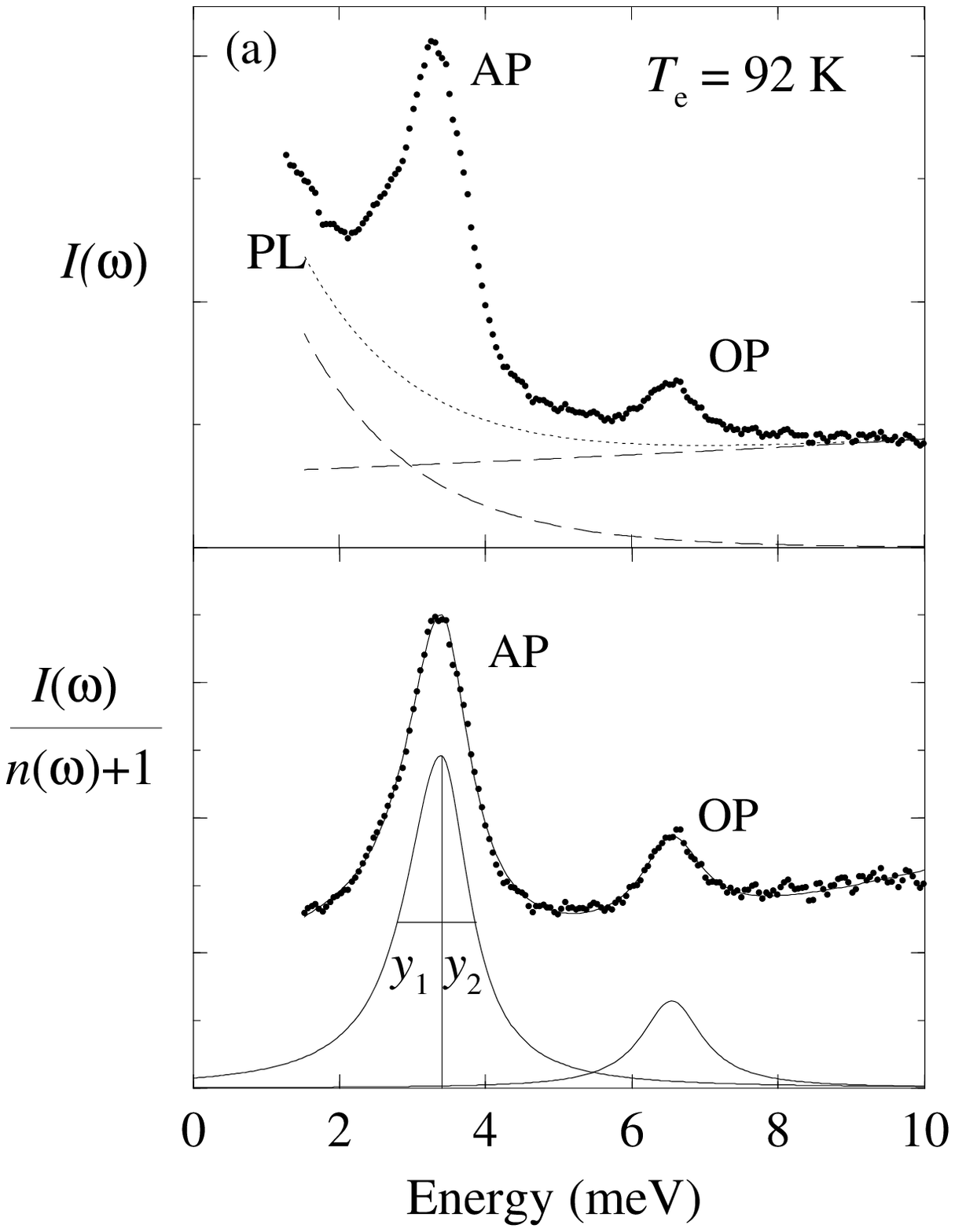}
\end{center}
\begin{center}
\epsfxsize = 10 cm 
\epsfbox{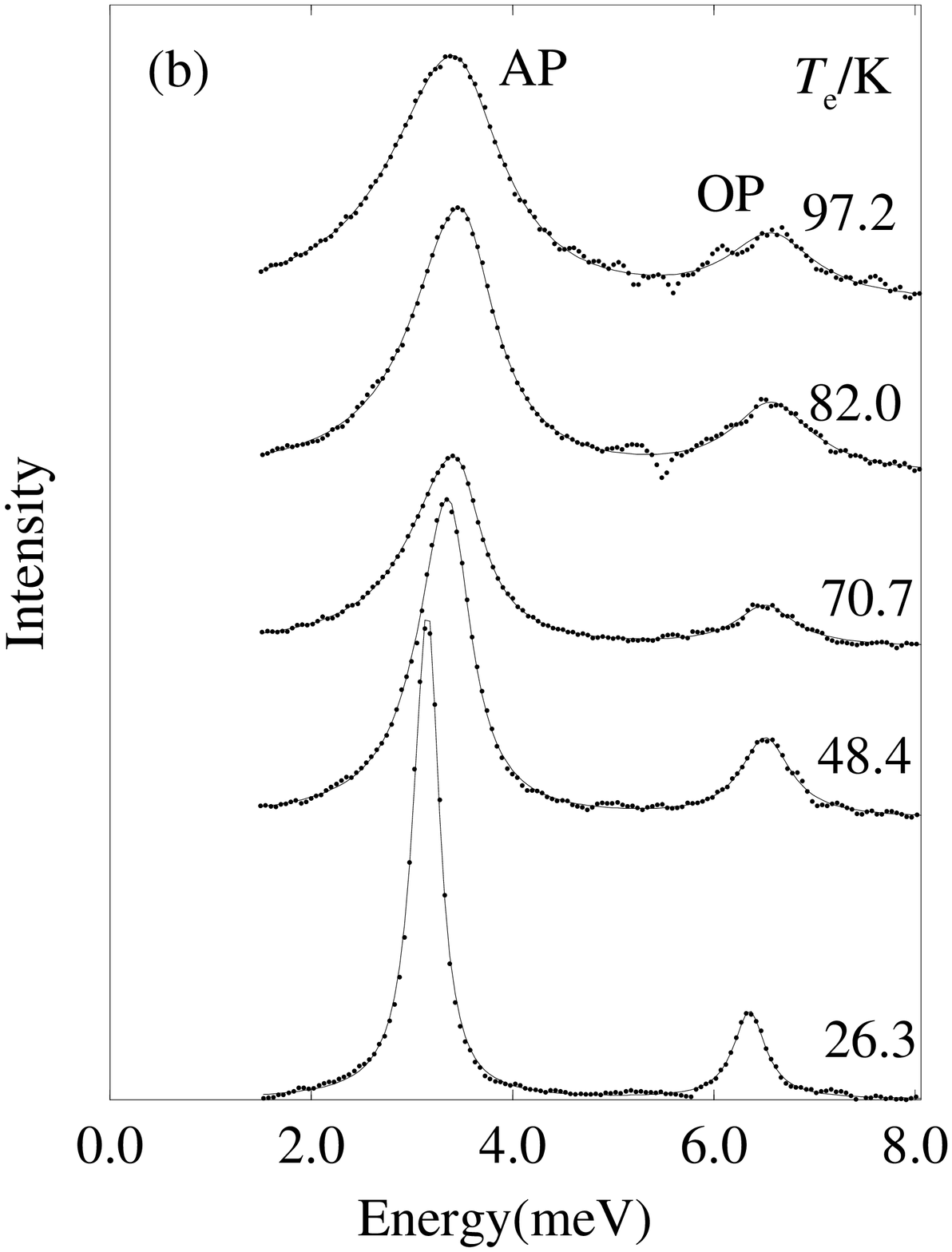}
\end{center}
\caption{(a) A spectrum $I(\omega)$ measured for an electron 
temperature $T_{\rm e} = 93~{\rm K}$ is shown in the upper panel and 
the spectrum corrected for the Raman occupation factor $(n(\omega)+1)$ 
is shown in the lower panel; $q = 1.6 \times 10^{5}~{\rm cm}^{-1}$.  
The points show the experimental spectrum; the solid lines indicate 
the fit to the spectrum (mainly obscured by the points due to the good 
fit) and the fit with the exponential backgrounds due to the bulk and 
hot PL removed.  In the upper panel the bulk and hot PL contributions 
are indicated by the dashed lines, and the total PL contribution by 
the dotted line.  The HWHMs (half-width at half maximum), $y_{1}$ and 
$y_{2}$, used to parameterise the peak widths, are indicated.  (b) 
Corrected Raman spectra (with the background PL spectra subtracted and 
the $(n(\omega)+1)$ occupation factor removed) of the acoustic (AP) 
and optic (AP) plasmons for various electron temperatures $T_{\rm e}$; 
$q = 1.6 \times 10^{5}~{\rm cm}^{-1}$.  Note the broadening and 
developing asymmetry of the AP with increasing $T_{\rm e}$, resulting 
from Landau damping.}
\label{Tdepstack}
\end{figure}

A Raman spectrum for $T_{\rm e} = 93$~K is shown in Fig.  
\ref{Tdepstack}(a); at such high electron temperatures the Raman 
signals are superimposed on quite intense background PL arising from 
the quantum well inter-band transition responsible for the Raman 
resonance (`hot' PL, at small Raman shifts) and also from the GaAs 
buffer layer (bulk PL, at high Raman shifts).  In addition, the 
importance at such elevated temperatures of the occupation factor in 
the Raman intensity ($n(\omega)+1$ in Eq.  \ref{Raman}) is also 
responsible in part for the large intensity at small Raman shifts.  In 
order to facilitate comparison between experiment and theory, 
especially for the AP damping as characterised by Raman peak widths, 
the experimental spectra $I(\omega)$ were first divided by the 
occupation factor $(n(\omega)+1)$, as shown in Fig.  
\ref{Tdepstack}(a).  Fits to the lineshapes $I(\omega)/(n(\omega)+1)$ 
were then obtained using exponential tails to describe the PL 
contributions to the experimental lineshapes; the result of such a fit 
is indicated in Fig.  \ref{Tdepstack}(a).  The exponential bulk and 
hot PL tails were then removed to give a corrected Raman signal, also 
shown in Fig.  \ref{Tdepstack}(a), in which the effect of the 
occupation factor $(n(\omega)+1)$ has been removed.

Fig.  \ref{Tdepstack}(b) shows Raman spectra measured for a range
of electron temperatures $T_{\rm e}$, for an in-plane wavevector
transfer $q = 1.6 \times 10^{5}~{\rm cm}^{-1}$.  These spectra have
been corrected as described above to leave signals due principally to
the plasmons.  Two peaks are present, which we assign to the acoustic
and optic plasmons of the 2DEG.\cite{Kainth98} The OP is seen to
broaden symmetrically as $T_{\rm e}$ rises, whereas the
AP is seen to broaden asymmetrically and to a much greater extent.

We ascribe the asymmetric broadening of the AP to Landau damping
arising from the increasing interaction with the lower lying SPE
continuum as $T_{\rm e}$ is increased (as discussed in Section
\ref{Ramanth} and illustrated in Fig.  \ref{ThLDspectra}).  The
broadening with $T_{\rm e}$ of the OP peak may be attributable to
increased scattering by unionised impurities in the sample bulk as the
temperature increases; this scattering is greater for the OP than the
AP as away from the bilayer region the electric fields associated with
the symmetric OP are much greater than those of the antisymmetric
AP.\cite{Fasol89} We have also shown that this effect can account for
the observed dependence on 2DEG mobility of the relative strengths of
Raman scattering by the AP and OP.\cite{Kainth99}

The observed AP width involves several contributions --- Gaussian
terms such as inhomogeneous broadening effects, spectrometer
resolution, and broadening due to the spread of in-plane wavevectors
accepted by the numerical aperture of the collecting lens, and
Lorentzian contributions from electron scattering (described by the
parameter $\gamma$ in Eq.  \ref{Lindhard}) and Landau damping.  These
combine according to:
\begin{equation}
\Delta\omega_{\rm L}\Delta\omega_{\rm tot}+(\Delta\omega_{\rm G})^{2}=
\Delta\omega_{\rm tot}^{2}\ \ ,
\label{Dobryakov}
\end{equation}
\noindent where $\Delta\omega_{\rm L}(\Delta\omega_{\rm G})$ is the
Lorentzian (Gaussian) contribution to the total half-width at half
maximum (HWHM), $\Delta\omega_{\rm tot}$.\cite{Dobryakov69} To allow
easy comparisons with the calculated spectra, the Gaussian
contributions to the half-widths were deconvolved from the
experimental spectra by adapting a method originally due to Dobryakov
and Lebedev.\cite{Dobryakov69} As a result the net Gaussian
contribution to the FWHM (full-width at half maximum) was estimated to
be 0.06 meV, in good agreement with the observed width of the Rayleigh
scattered laser line (\textit{cf.} the single particle relaxation FWHM
$2\gamma = 0.18$ meV).

Because the AP lineshape is asymmetric, HWHMs on either side were
measured and Eq.  \ref{Dobryakov} used to subtract $\Delta\omega_{\rm
G}$ in each case to produce $y_{1}$ and $y_{2}$, the HWHMs of the
Lorentzians on the low and high energy sides, respectively
(illustrated in Fig.  \ref{Tdepstack}(a)).  The width of the
AP, once Gaussian contributions have been subtracted, is a measure of
the total homogeneous damping experienced by the plasmon mode.  The
difference between $y_{1}$ and $y_{2}$, on the other hand, gives
information about the energy-dependent damping responsible for the
observed asymmetry of the AP peak; thus $y_{1} - y_{2}$ gives a direct
measure of the Landau damping experienced by the AP. In this way, by
comparing $y_{1}$ and $y_{2}$ with theoretically calculated
half-widths, we have a direct handle on the many-body interactions of
the AP with the SPE continuum.

We should note that the SPE band itself is potentially observable in
both polarised and depolarised Raman spectra.  However, at low
temperatures no SPE signal was observed in either polarisation for $q
= 1.6 \times 10^{5}~{\rm cm}^{-1}$.  For $q = 1.16 \times 10^{5}~{\rm
cm}^{-1}$ the SPE band is in a spectral region too close (a Raman
shift less than $\sim 1$ meV) to the exciting laser line to be
observed.

\subsection{Total AP damping}

\begin{figure}
\begin{center}
\epsfxsize = 12 cm 
\epsfbox{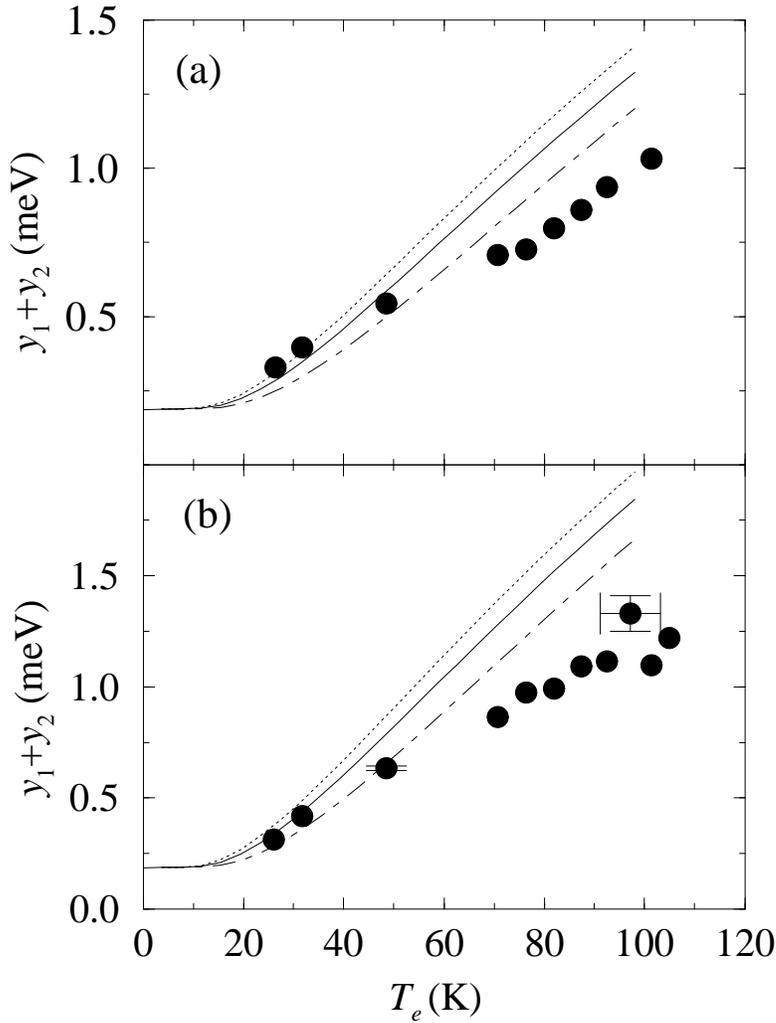}
\end{center}
\caption{Experimental ($\bullet$) and theoretical (curves) results for
the total width of the AP as a function of $T_{\rm e}$.  Calculations
have been performed within the RPA (dashed line), Hubbard
approximation (dotted line) and finite temperature STLS (solid line).
(a) $q = 1.16 \times 10^{5}~{\rm cm}^{-1}$, (b) $q = 1.6\times
10^{5}~{\rm cm}^{-1}$.}
\label{Totwdth}
\end{figure}

We now consider the comparison between the experimentally observed and
theoretically calculated AP total linewidths as a function of
temperature.  Fig.  \ref{Totwdth}(a) shows a plot of the
experimentally determined FWHM ($y_{1} + y_{2}$) for $q = 1.16 \times
10^{5}~{\rm cm}^{-1}$, after removing the Gaussian contributions as
described above.  The experimental uncertainties in these FWHM values
are of the order of $\pm~0.02$ meV for low $T_{\rm e}$, rising to
$\pm~0.06$ meV for higher temperatures.  Fig.  \ref{Totwdth}(b) shows
the corresponding results for $q = 1.6 \times 10^{5}~{\rm cm}^{-1}$.
The three curves show the results of calculations of the FWHM
determined within: (i) the RPA; (ii) the H-RPA; (iii) the finite
temperature STLS formalism described in Section \ref{Theorysec}.

We can see from Fig.  \ref{Totwdth} that the larger the magnitude of
the local field correction $G_{11}$, the greater the increase in
damping with increasing $T_{\rm e}$; as the Hubbard intra-layer local
field correction $G_{11}$ is greater than that obtained within the
STLS, for a given $T_{\rm e}$ the H-RPA predicts a
greater width than that calculated using the STLS approximation, which
is in turn greater than that obtained within the RPA.

For low $T_{\rm e}$, for both wavevectors, the experimentally
determined FWHM and the calculated values agree quite well, especially
as it should be noted that no free fit parameters were used for these
curves --- the 0 K theoretical FWHM was fixed from the experimental
single particle relaxation times (Section \ref{SdHsec}).  But since
the experimental uncertainties at these values of $T_{\rm e}$ are
smaller than the point size in the Figure, the better agreement
between the data and the H-RPA and STLS curves, as opposed
to the RPA calculation, is significant, and suggests that it is
necessary to include exchange-correlation effects even for these
densities.

However, the experimentally determined plasmon widths increase with
$T_{\rm e}$ much more slowly than is predicted theoretically, and for
$T_{\rm e} \gtrsim T_{\rm F}$ drop below even the values predicted by
the RPA. This could be interpreted in terms of a reduction in the
significance of correlation effects at these temperatures, at which
the RPA then provides a better description.  However, the STLS
calculation, which should take full account of the effects of non-zero
temperature, does not show this behaviour, reflecting the relative
constancy of the intra-layer local field correction $G_{11}$ over this
temperature range (see Fig.  \ref{GvsT}).  So it appears that for
$T_{\rm e} \gtrsim T_{\rm F}$, further theoretical developments are
necessary.

These calculations take no account of any dependence on temperature of
the single particle relaxation rate $\gamma$.  However, if anything we
would expect the single particle lifetime to decrease with increasing
temperature, leading to even larger theoretical linewidths than those
calculated here and increasing yet further the discrepancy between
experiment and theory.

\subsection{AP asymmetry}

\begin{figure}
\begin{center}
\epsfxsize = 12 cm 
\epsfbox{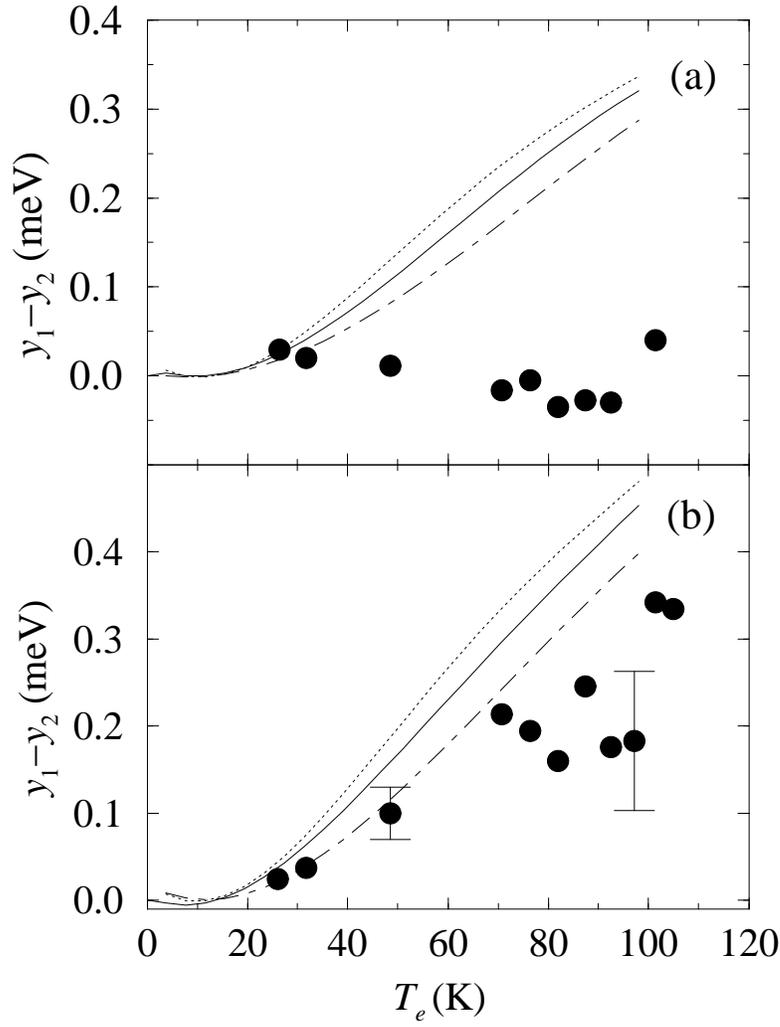}
\end{center}
\caption{The asymmetry of the AP (defined as $y_{1}-y_{2}$), as for
Fig. \ref{Totwdth}.}
\label{Diffwidth}
\end{figure}

For a given $T_{\rm e}$ it is expected that the low energy side of the
AP peak will be more affected by Landau damping than the high energy
side, as the SPE continuum lies below the AP dispersion curve.  A
measure of this asymmetric broadening is $(y_{1}-y_{2})$, which is
plotted in Fig.  \ref{Diffwidth} for both experiment and theory, again
for both $q = 1.16$ and $1.6 \times 10^{5}~{\rm cm}^{-1}$.

It is expected that $(y_{1} - y_{2}) \rightarrow 0$ as $T_{e} 
\rightarrow 0$ (the peak shape would be a symmetric Lorentzian) and 
increase with increasing $T_{\rm e}$.  Experimentally, for $q = 1.6 
\times 10^{5}~{\rm cm}^{-1}$ (Fig.  \ref{Diffwidth} (b)), 
$(y_{1}-y_{2})$ increases with low $T_{\rm e}$ as expected, but at 
higher temperatures ($T_{\rm e} \sim 70$~K) it becomes roughly 
constant.  For the smaller wavevector of $q = 1.16 \times 10^{5}~{\rm 
cm}^{-1}$ (Fig.  \ref{Diffwidth} (a)), the discrepancy between 
experiment and theory is greater and $(y_{1}-y_{2})$ is essentially 
zero over the measured temperature range.  In contrast to this, all 
the theoretical models predict a continuous increase in 
$(y_{1}-y_{2})$ with rising $T_{\rm e}$.  It is important to note that 
any temperature dependence of the single particle lifetime ($\sim 
1/2\gamma$) is unlikely to affect this measure of the asymmetry of the 
peak; we are sensitive here only to the effects of Landau damping.

\begin{figure}
\begin{center}
\epsfxsize = 12 cm 
\epsfbox{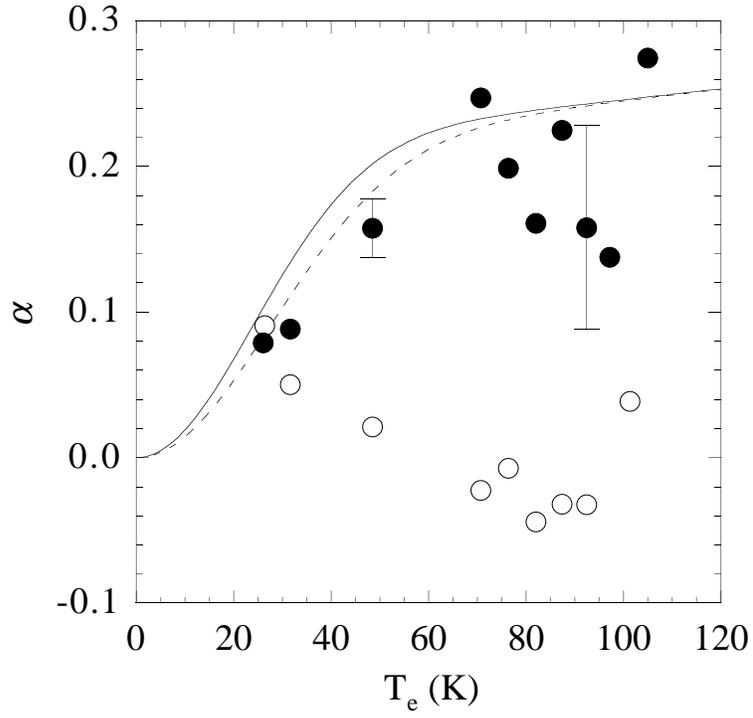}
\end{center}
\caption{The degree of asymmetry of the AP Raman peak, $\alpha =
(y_{1}-y_{2})/(y_{1}+y_{2})$, for experiment (points) and theory,
within the finite temperature STLS formalism (curves), for wavevectors
$q = 1.16$ ($\circ$, dashed line) and $q = 1.6 \times 10^{5}~{\rm
cm}^{-1}$ ($\bullet$, solid line).}
\label{DofA}
\end{figure}

Another measure of the Landau damping is provided by the degree of 
asymmetry, $\alpha=(y_{1}-y_{2})/(y_{1}+y_{2})$, shown in Fig.  
\ref{DofA} for experiment and for our STLS calculation.  For the two 
wavectors considered, the STLS theory predicts essentially the same 
variation with temperature, with $\alpha$ levelling off near $T_{\rm 
F}$.  This insensitivity of $\alpha$ to the wavevector is to be 
expected, as both the AP and the SPE continuum have a similar, 
approximately linear, dependence on $q$, for the small wavevectors ($q 
\ll k_{\rm F}$) considered here.  However, the experimental results 
for the two wavevectors are markedly different.  For $q = 1.6 \times 
10^{5}~{\rm cm}^{-1}$ there is a fairly good agreement between 
experiment and theory over the whole temperature range, although the 
experimental uncertainties are large ($\sim \pm~40 \%$ for high
$T_{\rm e}$); note, though, that the calculations overestimate the
FWHM of the AP peak at higher temperatures.  For $q = 1.16 \times
10^{5}~{\rm cm}^{-1}$ there is a significant discrepancy between
experiment and theory.

\subsection{AP energy}

\begin{figure}
\begin{center}
\epsfxsize = 12 cm 
\epsfbox{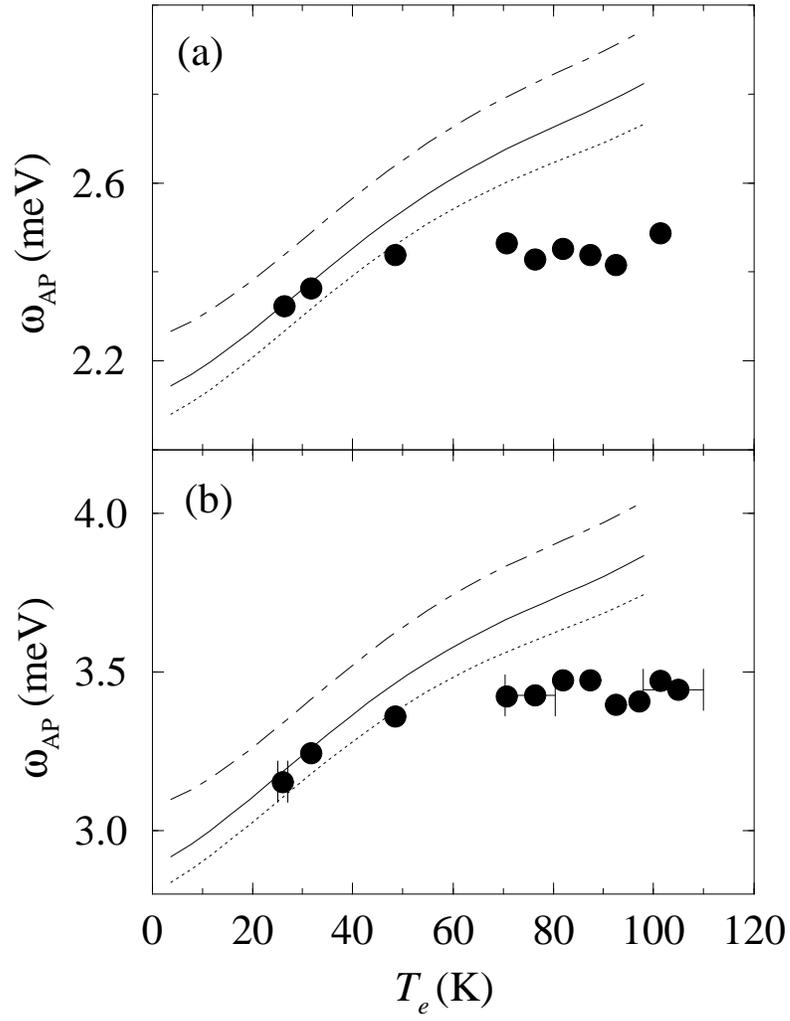}
\end{center}
\caption{AP Energy $\omega_{\rm AP}$, as for Fig.  \ref{Totwdth}.}
\label{APenergy}
\end{figure}

Figs.  \ref{APenergy} (a) and (b) show the energy $\omega_{\rm AP}$ of
the acoustic plasmon as a function of $T_{\rm e}$ for $q = 1.16$ and
$1.6 \times 10^{5}~{\rm cm}^{-1}$, for both experiment and theory.
The 2DEG density for the calculations was set so that the AP
dispersion obtained within the STLS approximation matched the
experimentally determined dispersion at low
temperature.\cite{Kainth99} (Previously,\cite{Kainth98} the density
was obtained using a fit determined within the H-RPA,
resulting in a slightly higher density than that used here; this
results in the slight shift of the RPA and H-RPA curves with respect
to our data for $(y_{1}+y_{2})$ (Fig.  \ref{Totwdth}) and
$(y_{1}-y_{2})$ (Fig.  \ref{Diffwidth}), compared with those presented
in Fig.  4 of Ref.  \onlinecite{Kainth98}.)

All the theoretical curves show a large increase ($\sim 30\%$) of the
AP energy with temperature, for both wavevectors.  The OP energy
$\omega_{\rm OP}$ was also found to increase with $T_{\rm e}$,
although by only $\sim 10\%$.  Such behaviour is well known for
classical Maxwell-Boltzmann plasmas,\cite{classical} and similar
calculations carried out for a 2D plasma correspond closely to our RPA
results at higher temperatures.

Fig.  \ref{APenergy} shows that the inclusion of exchange-correlation
effects reduces the calculated values of $\omega_{\rm AP}$, as would
be expected from the softening of the effective interaction potentials
$\widetilde{v}_{ij}$ (Eq.  \ref{eq:linear response3}).  The weak
temperature dependence of the STLS local field correction (see Fig.
\ref{GvsT}) does not significantly alter the variation of the plasmon
energy with $T_{\rm e}$.

Experimentally, for $T_{\rm e} \lesssim 50$ K ($T_{\rm F} = 78$ K),
$\omega_{\rm AP}$ increases with $T_{\rm e}$, in keeping with the
calculations.  However, for both wavevectors, for $T_{\rm e} \gtrsim
50$ K, $\omega_{\rm AP}$ remains approximately constant, in contrast
with the theories, all of which predict a steady increase in the
plasmon energy.  The variation of the OP energy $\omega_{\rm OP}$ with
temperature follows essentially the same pattern as that of the AP
(see Fig.  \ref{Tdepstack}(b)), although there is better agreement
between experiment and theory due to the smaller predicted increase of
$\omega_{\rm OP}$ ($\sim 10\%$) with $T_{\rm e}$.

\subsection{Discussion}


Above $T_{\rm e} \sim 50$ K, there are significant discrepancies
between experiment and theory. All three calculations show an increase in
$\omega_{\rm AP}$, contrary to the observed plateau behaviour, and all three
overestimate the width and asymmetry.  Not only does the RPA fail to
describe our results, but so does the full temperature dependent STLS
scheme. We believe that these discrepancies 
are significant, suggesting a failure of the static local field theories 
used here to describe adequately the excitation spectra of 
non-degenerate low-dimensional electron gases.

\begin{figure}
\begin{center}
\epsfxsize = 10 cm 
\epsfbox{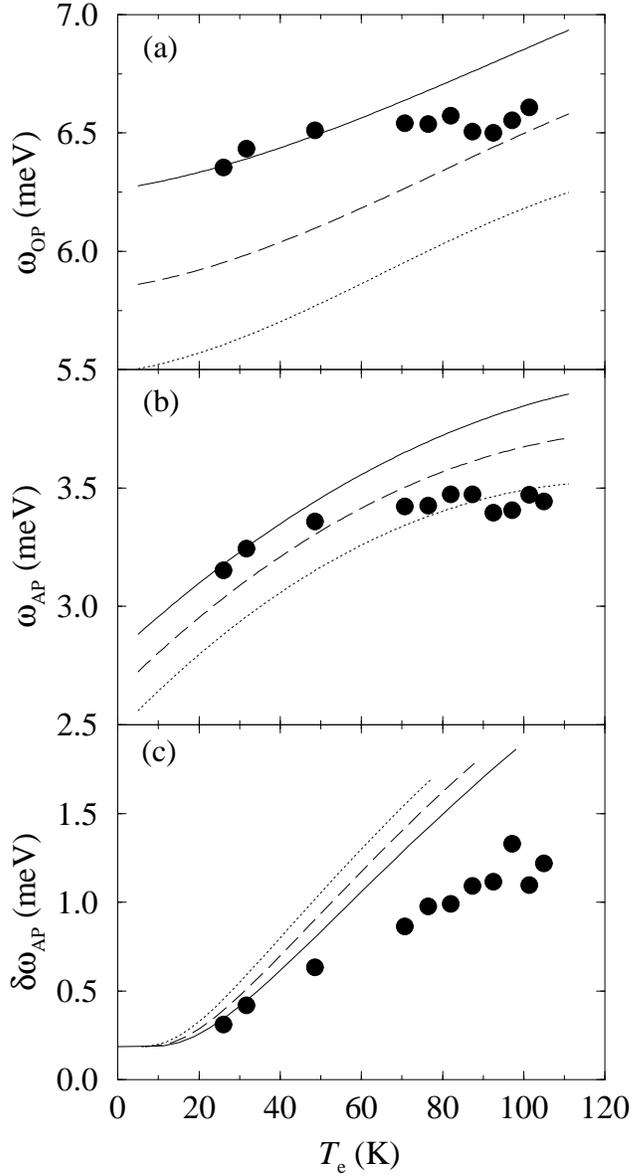}
\end{center}
\caption{Experimental ($\bullet$) and theroretical (curves)
temperature ($T_{e}$) dependence for $q = 1.6 \times 10^{5}~{\rm
cm}^{-1}$ of (a) the OP energy $\omega_{\rm OP}$; (b) the AP energy
$\omega_{\rm AP}$; (c) the AP damping $\delta\omega_{\rm AP} \equiv
(y_{1}+y_{2})$.  The theoretical curves have been determined for
electron densities of (solid lines) 1.95, (dashed lines) 1.7 and
(dotted lines) $1.5 \times 10^{11}~{\rm cm}^{-2}$, using STLS local
field factors given in Fig.  \ref{GvsT}.}
\label{densdep}
\end{figure}

To demonstrate that this is the case,
we must firstly
to examine whether alternative causes of the observed plasmon behaviour, 
such as changes with $T_{e}$ of
the 2DEG densities, or of their background screening, could explain our
experimental results.
Fig. \ref{densdep} explores the effect of possible changes in the 2DEG
densities by plotting, for various 2DEG densities, calculations of
$\omega_{\rm OP}$, $\omega_{\rm AP}$ and the AP FWHM
$\delta\omega_{\rm AP} \equiv (y_{1}+y_{2})$ as functions of $T_{e}$
for $q = 1.6 \times 10^{5}~{\rm cm}^{-1}$, together with the
corresponding experimental results.  (For expediency, the STLS local
field factors self-consistently determined for a density of $1.95
\times 10^{11}~{\rm cm}^{-2}$ (Fig.  \ref{GvsT}) were used in the
calculations for all densities; this can be justified by the small
variation in the low $T_{e}$ values for $G_{11}$ and $G_{12}$ over the
relatively small range of densities considered.)  The observed
temperature dependence of $\omega_{\rm AP}$ (Fig.  \ref{densdep}(b))
would require a significant reduction in 2DEG density (by about 30\%
for the highest measurement temperatures), though a much smaller
reduction ($\sim 15\%$) would account for the behaviour of
$\omega_{\rm OP}$ (Fig.  \ref{densdep}(a)).  Moreover, Fig.
\ref{densdep}(c) shows that the AP damping $\delta\omega_{\rm AP}$
actually increases with decreasing density, worsening the discrepancy
between experiment and theory; for a given $T_{\rm e}$, the smaller
$T_{\rm F}$ resulting from a decrease in density effectively increases
the reduced temperature $T_{\rm e}/T_{\rm F}$ and enhances the Landau
damping.  So although a reduction in density with temperature could
account for the observed $T_{\rm e}$-dependence of $\omega_{\rm AP}$,
the corresponding temperature dependence of $\omega_{\rm OP}$ and
$\delta\omega_{\rm AP}$ cannot be reconciled to this.

\begin{figure}
\begin{center}
\epsfxsize = 10 cm 
\epsfbox{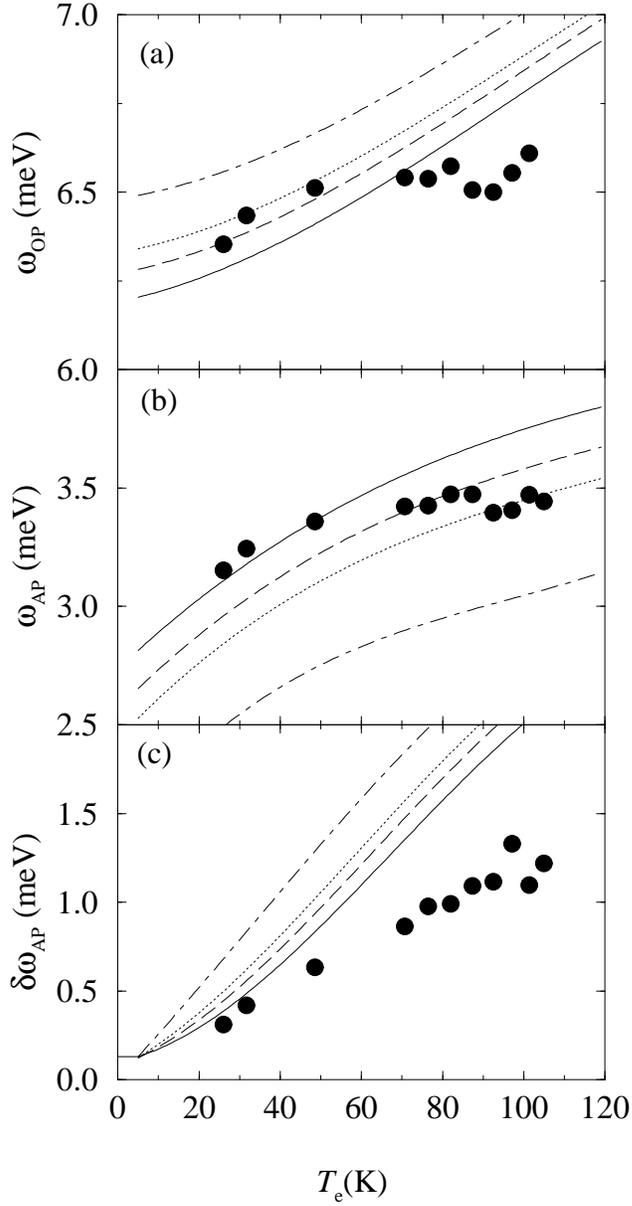}
\end{center}
\caption{Experimental ($\bullet$) and theroretical (curves)
temperature ($T_{e}$) dependence for $q = 1.6 \times 10^{5}~{\rm
cm}^{-1}$ of (a) the OP energy $\omega_{\rm OP}$; (b) the AP energy
$\omega_{\rm AP}$; (c) the AP damping $\delta\omega_{\rm AP} \equiv
(y_{1}+y_{2})$.  The theoretical curves have been determined within
the RPA using a Hubbard local field correction for a total electron
density $N_{1} + N_{2} = 3.9 \times 10^{11}~{\rm cm}^{-2}$ with $N_{1}
/ N_{2} = 1.0$ (solid lines), 1.9 (dashed lines), 2.3 (dotted lines)
and 4.0 (dot-dashed lines).}
\label{unequal}
\end{figure}

We have assumed throughout that the two 2DEGs are of equal density, an
assumption borne out at low temperatures from the form of the plasmon
dispersions.\cite{Kainth98,Kainth99} However, a temperature induced
asymmetry in the electron distribution between the two quantum wells
(the total electron density remaining constant) might occur;
calculations (within the H-RPA for computational
simplicity) show (Fig.  \ref{unequal} (b)) that this would result in a
reduction of $\omega_{\rm AP}$, and that agreement between experiment
and theory is possible at the highest measurement temperatures for
$N_{1}/N_{2} \sim 3$ (i.e. $N_{1} \sim 2.9$ and $N_{2} \sim 1.0 \times
10^{11}~{\rm cm}^{-2}$).  However, not only is this a very significant
asymmetry, but again neither the $T_{\rm e}$-dependence of
$\omega_{\rm OP}$, or that of $\delta\omega_{\rm AP}$, fall into line
with this.  An asymmetric density distribution results in a slight
increase in $\omega_{\rm OP}$ (Fig.  \ref{unequal} (a)) as the two
plasmon modes decouple and the OP becomes more localised in the higher
density layer; at the same time the AP becomes more localised in the
lower density layer, leading to the observed reduction in energy.  An
increasing density asymmetry also leads to an increase in the Landau
damping of the AP, which comes closer to (and enters, for a large
difference between the densities) the SPE continuum of the higher
density layer.\cite{Liu96}

The variations of $\omega_{\rm OP}$ and $\omega_{\rm AP}$ with $T_{\rm
e}$ (see Figs.  \ref{densdep} and \ref{unequal}) also indicate that
changes in background screening (from, \textit{e.g.}, un-ionised
donors in the doped AlGaAs layers) are not responsible for the
reduction in $\omega_{\rm AP}$ from the theoretical values.  If this
were the case, we would expect $\omega_{\rm OP}$ to be affected to a
much greater extent, as seen in the widths of the two plasmon peaks
(see Section \ref{Res1} and Fig.  \ref{Tdepstack}).  In addition,
although increased screening would reduce $\omega_{\rm AP}$, the SPE
continuum would be unaffected, resulting in increased Landau damping,
contrary to observation.

Therefore it appears that the inability of our calculations to model
our experimental results for the energy and damping of the AP cannot
be simply attributed to an experimental artifact, such as a decrease
in 2DEG density with temperature. We must thus examine whether
the use of a static local field correction provides  
an accurate description of the effect of   
exchange and correlation on
the excitation spectra of
non-degenerate quasi-two-dimensional electron systems.
In order to obtain better 
agreement with experiment for the AP energy, the 
magnitude of $G_{11}$ must increase with temperature, whereas to 
obtain an improved fit for the measured AP damping, $G_{11}$ must 
decrease. \cite{footnote}
 This contradiction suggests that a real local field factor,
(as calculated within the framework of a static theory) cannot 
explain our observed results. This is one of the central conclusions of this
work.



\begin{figure}
\begin{center}
\epsfxsize = 12 cm 
\epsfbox{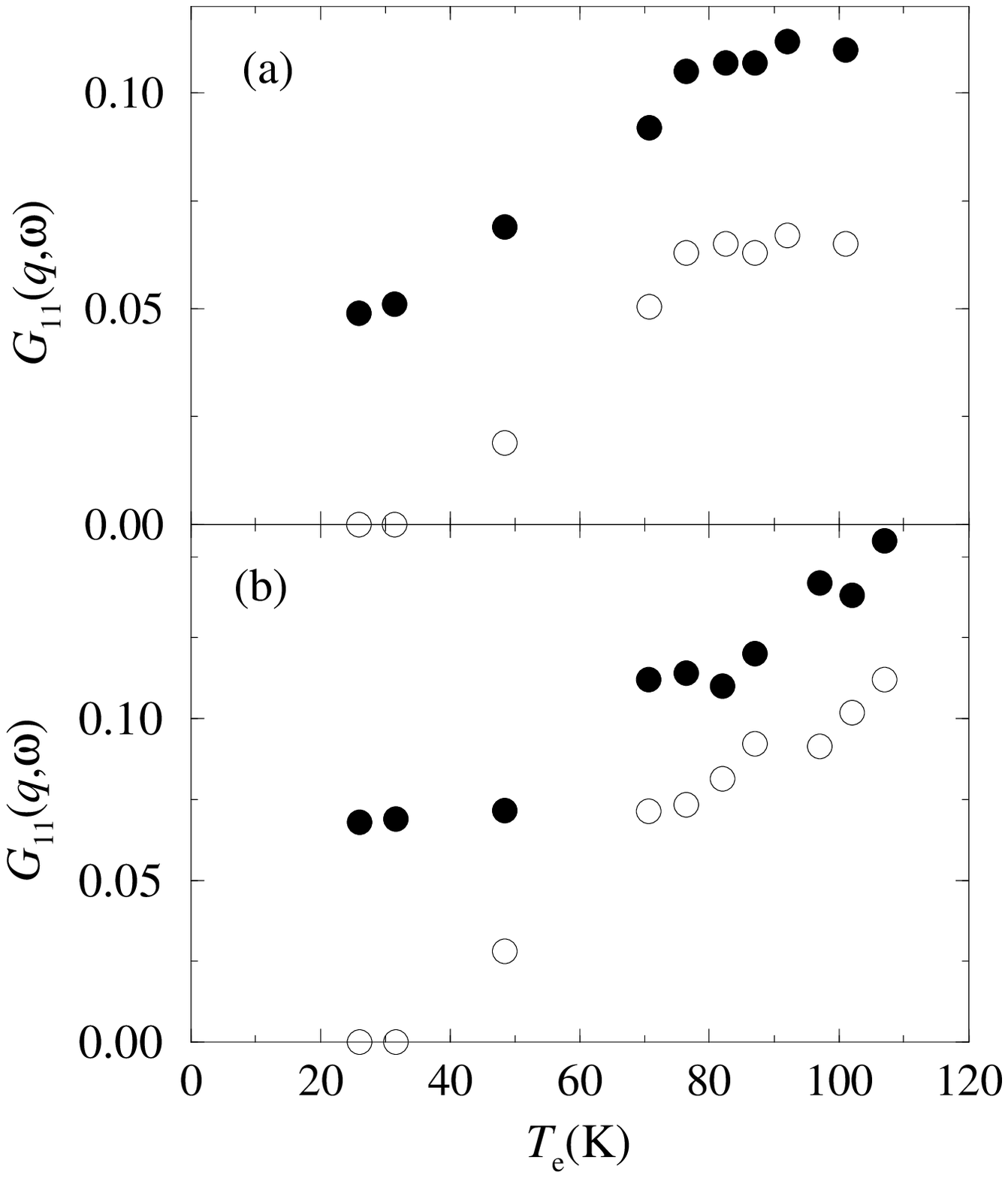}
\end{center}
\caption{The temperature ($T_{e}$) dependence of the real part 
$G'_{11}$ ($\bullet$) and imaginary part $G''_{11}$ ($\circ$) of the 
intra-layer complex local field factor $G_{11}(q,\omega)$, used as a 
fitting parameter to provide agreement between experiment and theory 
for the AP energy $\omega_{\rm AP}$ and the AP damping 
$\delta\omega_{\rm AP}$.  The inter-layer local field factor is set to 
$G_{12} = 0$ (a valid approximation given its relatively small 
magnitude).  (a) $q = 1.16 \times 10^{5}~{\rm cm}^{-1}$, (b) $q = 
1.6\times 10^{5}~{\rm cm}^{-1}$.}
\label{complexG}
\end{figure}

We are led therefore to consider whether a complex local field correction,
$G_{11} = G'_{11} + i G''_{11}$, as suggested 
in the context of a dynamic local field 
theory \cite{Mougdil95,Richardson94} is any more successful in characterising
the observed behaviour. 
For a given temperature 
$T_{\rm e}$, $G'_{11}$ (the real part of $G_{11}$) was determined from 
the energy of the acoustic plasmon $\omega_{\rm AP}$; $G''_{11}$ (the 
imaginary part of $G_{11}$) was found to have a negligible affect on 
$\omega_{\rm AP}$.  Singwi \textit{et al.}\cite{STLS} have argued that 
the imaginary part of the local field correction is important in 
determining the lifetimes of plasmon modes.  So, using the derived 
value for $G'_{11}$, the imaginary part $G''_{11}$ was then determined 
by fitting the plasmon damping $\delta\omega_{\rm AP}$.  In order to 
give a reduction in the AP damping, to bring theory in line with 
experiment, $G''_{11}$ was determined to be positive.  As well as now 
providing complete agreement between experiment and theory for the 
$T_{\rm e}$-dependence of $\omega_{\rm AP}$ and $\delta\omega_{\rm 
AP}$, the resulting complex local field factor, $G_{11} = G'_{11} + i 
G''_{11}$, was also found to provide a more accurate description of 
the optic plasmon energy $\omega_{\rm OP}$ and the AP asymmetry for 
both wavevectors $q$ considered here.  The variation with temperature 
$T_{\rm e}$ of these fitting parameters, the real and imaginary parts 
of $G_{11}$, is shown in Fig.  \ref{complexG} for the two wavevectors 
$q$.  For $T_{\rm e} \lesssim T_{\rm F}/2$, the real and imaginary 
parts of the local field correction are similar to the values obtained 
within the STLS; in particular $G''_{11}$ is zero.  However, for 
$T_{\rm e} \gtrsim T_{\rm F}/2$, the temperature range over which the 
static STLS approach was found to fail, we see that both the real and 
imaginary parts of the intra-layer local fields increase with $T_{\rm 
e}$.  This is perhaps surprising, given our initial expectation that 
exchange-correlation effects should become less important with 
increasing temperature.  Nevertheless, it is significant that the only 
way in which we have been able to obtain agreement between experiment 
and theory is through the incorporation of a complex local field 
correction.

Finally, we observe that our conclusions are consistent with the 
results of measurements of the AP-mediated Coulomb drag between two 
electron layers in samples similar to that studied 
here.\cite{Hill97,Noh98} In particular, in such experiments the onset 
temperature for plasmon enhancement of the drag rate is determined by 
the energies of the acoustic plasmons.  A suppression of the AP 
energies from that expected within the RPA, as observed in our Raman 
measurements, should lead to the plasmon enhancement in the Coulomb 
drag rate occurring at a lower temperature than that predicted by the 
RPA; this is indeed observed in Refs.  \onlinecite{Hill97} and 
\onlinecite{Noh98}.  On the other hand, an increase in the damping of 
the plasmons leads to an increase in the magnitude of the plasmon 
enhancement in the drag rate.  Hill \textit{et al.} found that 
although the use of a Hubbard static local field factor resulted in 
better agreement between theory and experiment for the onset 
temperature, due to the suppression of the AP energies, the resultant 
increase in the plasmon damping leads to an overestimate of the 
magnitude of the enhancement,\cite{Hill97} which mirrors the results 
of the Raman measurements of the AP damping presented here 
(\textit{i.e.} at higher temperatures we find the plasmon damping is 
reduced from that predicted by theory).  Our use of a complex local 
field factor to reconcile the behaviour of the plasmon energy and 
damping is consistent with the conclusions of Noh \textit{et al.}, who 
suggested that a dynamic local field theory could account for their 
Coulomb drag results,\cite{Noh98} although they implied that this 
would lead to a broadening of the plasmon modes, whereas we find the 
damping (as determined from the Raman lineshape) to be less than that 
predicted by the RPA.

\section{Summary}

To conclude, we have analysed the temperature dependence of the Landau 
damping of the acoustic plasmon in an electron bilayer, with a view to 
obtaining a measure of the importance of exchange-correlation effects 
in such systems.  To this end, we have modelled the density response 
of our system using the Singwi-Tosi-Land-Sj\"{o}lander (STLS) 
approximation, in which we have taken full account of the effects of 
finite temperature.  The most significant message of these 
calculations is that non-zero temperature correlations within the STLS 
differ little from their zero temperature counterparts at low 
wavevector.  Increasing the temperature leads to an increase in the 
kinetic energy of the electron gas, implying a decrease in the 
significance of many body effects in determining its behaviour.  
Na\"{\i}vely we could have expected our temperature dependent local 
field corrections to show a transition from the H-RPA (a model 
including exchange-correlation effects to some degree) to the RPA (a 
mode in which exchange and correlation are not 
included).\cite{Kainth98} If this were the case we would have expected 
a much greater decline in the local field corrections towards zero as 
$T_{\rm e}$ is increased.  However, this does not seem to be the case 
for the static local field corrections we have calculated within the 
STLS.

Experimentally, we determined the electron temperature of the 2DEGs
from measurements of the photoluminescence, which gives an accurate
determination of the temperature under the same conditions as the
Raman measurements.  Our Raman lineshapes displayed the clear
signature of Landau damping of the acoustic plasmon, indicated by the
asymmetric form of the Raman peak.  Careful analysis of the spectra
allowed us to extract three parameters which characterise the
lineshape and energy of the acoustic plasmon: $y_{1}$ and $y_{2}$, the
low and high Raman shift half widths of the plasmons, and the AP
energy itself.  These parameters were compared with theoretically
calculated values, determined within the RPA and within the H-RPA
and STLS local field theories, thereby giving us a means to analyse
the success of these theories in modelling electron gases of
intermediate degeneracy.  We found that for $T_{\rm e} \lesssim T_{\rm
F}/2$ the inclusion of exchange-correlation effects using a local
field correction is necessary --- indeed sufficient --- to obtain good
quantitative agreement for the asymmetric Landau damping.  Conversely,
for $T_{\rm e} \gtrsim T_{\rm F}/2$ none of the theories modelled
successfully the AP asymmetry, or its upward shift in energy with
increasing $T_{\rm e}$.  Further, the behaviour of the AP energy and
damping cannot be reconciled to the calculations by a reduction with
$T_{\rm e}$ of the 2DEG density, or a redistribution of electrons
between the two quantum wells.  This leaves only exchange-correlation
effects as the cause of the
discrepancy between experiment and theory.  We find that the STLS
approach and indeed any static local field theory cannot effectively
model the exchange-correlation effects apparent in the experimental
data. 

We should note also that we observed a greater discrepancy between
experiment and theory for a wavevector of $q \sim 0.1 k_{\rm F}$ than
for $q \sim 0.14 k_{\rm F}$.  Indeed, for the lower wavevector the
Landau damping of the AP appeared to be effectively constant over the
whole temperature range considered.

There are a number of alternative, although less widely applied,
theoretical approaches to that of the STLS. 
It may be that a non-local scheme
(see, \textit{e.g.}, Refs.  \onlinecite{Ryan} and \onlinecite{Luo93})
may be necessary for a true description of the excitation spectra of
low-dimensional electron systems.  However, from our work the most promising
candidate is probably a quantum STLS approach, employing a frequency
dependent local field correction,\cite{Mougdil95,Schweng93} which may
go towards providing a more complete agreement between experiment and
theory; we have shown that we are able to obtain excellent agreement
between experiment and theory using as a fitting parameter a complex
local field factor, as suggested in the context of a dynamical local
field theory.  Our work shows that the temperature dependence of
Landau damping is likely to be a valuable experimental probe of these
theories.

\section{Acknowledgements}

We thank the UK Engineering and Physical Sciences Research Council,
the Royal Society, and the UK Defence Evaluation and Research Agency
for support.  We are grateful to K. Zolleis for the Shubnikov-de Haas
measurements and to C.H.W. Barnes for numerical assistance.

\appendix
\section{}
\label{STLSapp}

We describe here our implementation of the STLS scheme for electron
bilayers.  The density response functions $\chi_{ij}(q,\omega)$ will
have poles on (or infinitesimally close to) the real $\omega$ axis
which correspond physically to the energies of the plasmon modes,
while the poles of the $\coth$ function in Eq.  \ref{eq:def_S(ij)}
will lie equispaced along the imaginary $\omega$ axis.  A Wick
rotation ($\omega \rightarrow {\rm{i}}\omega$) allows us to evaluate
the integral for the static structure factors $S_{ij}(q)$ in Eq.
\ref{eq:def_S(ij)} as a contour integral along the imaginary $\omega$
axis, where the poles of the integrand are now just the poles
($\gamma_{\nu}=2 \pi i\nu k_{\rm B} T/\hbar, \nu=0,\pm1,\pm2...$) of
$\coth (\hbar \omega/2 k_{\rm B}T)$.  The residue theorem then allows
a transformation of the integral into a sum over imaginary (Matsubara)
frequencies giving:
\begin{equation}
S_{ij}(q) = -\frac{k_{\rm B}T}{\sqrt{N_{i}N_{j}}}
\sum_{\nu=-\infty}^{\nu=\infty} {\chi_{ij}(q,\gamma_{\nu })} \ \ .
\label{eq:S_sum}
\end{equation}

For convenience, we now assume that the two layers are of equal
density, i.e. $N_{i}=N_{j}=N$.  Defining $\Phi_{ij}(q,\gamma_{\nu}) =
- (E_{\rm F}/N) \chi_{ij} (q,\gamma_{\nu})$, we have:
\begin{equation}
S_{ij}(q) =\sum_{\nu=-\infty}^{\nu=\infty}
{\Phi_{ij}(q,\gamma_{\nu})~\theta} \ \ ,
\label{eq:S_sum2}
\end{equation}
\noindent where $\theta = k_{\rm B}T/E_{\rm F} = T/T_{\rm F}$ and
$E_{\rm F}$ and $T_{\rm F}$ are the Fermi energy and temperature.
From Eq. \ref{eq:def_chi(ij)} we have:
\begin{equation}
\overline{\overline{\Phi}} = \frac{\Phi_{1}}{1 -
\widetilde{v}_{12}^{2}\Phi_{1}^{2}}~\left(
\begin{array}{cc}
1 & -\widetilde{v}_{12}\Phi_{1} \\
-\widetilde{v}_{12}\Phi_{1} & 1
\end{array}
\right)
\label{eq:def_phi(ij)}
\end{equation}
\begin{equation}
\Phi_{1} = \frac{\Phi_{\rm 2D}^{0}}{1 + \Phi_{\rm 2D}^{0}
\widetilde{v}_{11}} \ \ .
\label{eq:def_phi(i)}
\end{equation}
\noindent $\Phi_{\rm 2D}^{0}$ has been evaluated explicitly by
Phatisena \textit{et al.} who derived the form:\cite{Phatisena86}
\begin{equation}
\Phi_{\rm 2D}^{0}(q,\gamma_{\nu}) = \int_{0}^{\infty}{k~f(k)
\frac{\left|2\cos(\phi)\right|}{[(q^{4}/4 - k^{2}q^{2} -
\gamma_{\nu}^{2}) + q^{4}\gamma_{\nu}^{2}]^{1/4}}~{\rm d}k} \ \ ,
\label{eq:def_phi(0)1}
\end{equation}
\noindent where $f(k)$ is the Fermi-Dirac distribution function and
\begin{equation}
\tan(2\phi)=\frac{q^{2}\gamma_{\nu}^{2}} {q^{4}/4 - k^{2}q^{2} -
\gamma_{\nu}^{2}} \ \ .
\label{eq:def_phi(0)2}
\end{equation}
\noindent Note that in the case of a negative denominator $2\phi$ has
to be replaced by $\pi-2\phi$.

Thus in order to evaluate $S_{ij}(q)$ we must evaluate Eq.
\ref{eq:S_sum2} using Eq.  \ref{eq:def_phi(ij)}.  $S_{11}(q)$ can be
written as:
\begin{equation}
S_{11}(q) =
\sum_{\nu=-\infty}^{\infty}{\frac{\theta~\Phi_{1}(q,\gamma_{\nu})} {1
- \widetilde{v}_{12}^2(q) \Phi_{1}^{2}(q,\gamma_{\nu})}} \ \ ,
\label{S11sum1}
\end{equation}
\noindent which is reformulated as
\begin{equation}
S_{11}(q) = \underbrace{\sum_{\nu=-\infty}^{\infty}
{\theta~\Phi_{1}(q,\gamma_{\nu})}}_{S_{{ \rm{2D}}}(q)} +
\underbrace{\sum_{\nu=-\infty}^{\infty}
{\frac{\theta~\widetilde{v}_{12}^2(q) ~\Phi_{1}^{3}(q,\gamma_{\nu})}
{1 - \widetilde{v}_{12}^2(q)
~\Phi_{1}^{2}(q,\gamma_{\nu})}}}_{{\rm{\Delta S}}} \ \ .
\label{S11sum2}
\end{equation}
\noindent The first term in Eq.  \ref{S11sum2} is the structure factor
for a single electron layer, $S_{{\rm{2D}}}(q)$.  The second,
interlayer term $\Delta S$ is $O(\Phi_{1}^{3})$ and acts as a (small)
perturbation, reflecting the small extent to which the second layer
affects intra-layer correlations.

$S_{12}(q)$ is given by:
\begin{equation}
S_{12}(q) = -~\sum_{\nu=-\infty}^{\infty}
{\frac{\theta~\widetilde{v}_{12}^2(q) ~\Phi_{1}^{2 }(q,\gamma_{\nu})}
{1-\widetilde{v}_{12}^2(q)~\Phi_{1}^{2}(q,\gamma_{\nu})}}\ \ ,
\label{S12sum}
\end{equation}
\noindent which is $O(\Phi_{1}^{2})$ and hence again generally acts as
a perturbation.  The explicit evaluation of the sums to calculate
$S_{11}(q)$ and $S_{12}(q)$ is performed in an analogous manner to the
single layer case, as described by Bohm and Schweng.\cite{Schweng94}

We can note in passing that the success of the relatively simple H-RPA
approach when applied to bilayer systems can be traced directly to
Eqs.  \ref{S11sum2} and \ref{S12sum}: the H-RPA includes quite
effectively the $O(\Phi_{1})$ contribution to the local field in a
bilayer system.  The STLS formalism, however, includes the inter-layer
exchange-correlation and the perturbation of the intra-layer
correlations by the presence of the second layer.

We can rewrite Eq.  \ref{eq:def_G(ij)} for the local field factors
$G_{ij}(q)$ for the case of equal densities in terms of dimensionless
variables (${\bf Q} = {\bf q}/k_{F}$, ${\bf K} = {\bf k}/k_{F}$) as:
\begin{equation}
G_{ij}({\bf Q}) = \frac{1}{2 \pi} \int{{\rm d}{\bf K} \frac{{\bf Q}
\cdot {\bf K}}{|{\bf Q}||{\bf K}|} \frac{F_{ij}^{\prime}(|{\bf
K}|)}{F_{ij}^{\prime}(|{\bf Q}|)} ~[\delta_{ij} -
S_{ij}^{\prime}(|{\bf Q}-{\bf K}|)]}\ \ ,
\label{dimensionlessGij}
\end{equation}
\noindent where $S_{ij}^{\prime}(|{\bf Q} - {\bf K}|) = S_{ij}(|{\bf
q} - {\bf k}|)$ and $F_{ij}^{\prime}(|{\bf Q}|) = F_{ij}(|{\bf q}|)$.
Following Jonson in using the substitution ${\bf C} = {\bf Q} - {\bf
K}$, and with $a=C/Q$, we have as our final form:\cite{Jonson76}
\begin{eqnarray}
\nonumber G_{ij}({\bf Q}) &=& \frac{1}{2 \pi}\int_{0}^{\infty}{{\rm }d
C ~C~ [\delta_{ij} - S_{ij}^{\prime}(C)] } \\
&\times& \int_{0}^{2\pi} {{\frac{F_{ij}^{\prime}(|{\bf Q}|\sqrt{1 +
a^{2} + 2a\cos\phi})} {F_{ij}^{\prime}(|{\bf Q}|)}}
\frac{a\cos\phi}{\sqrt{1 + a^{2} + 2a\cos\phi}}} \ \ .
\label{final}
\end{eqnarray}

Eqs.  \ref{eq:S_sum2}, \ref{eq:def_phi(ij)} and \ref{final} were
solved self consistently using a method due originally to Ng to
accelerate (ensure) convergence.\cite{Ng74} For low number densities
(i.e. $r_{\rm s} \ge 2$) it was also found that the initial value of
$G_{11}(q)$ has to be chosen carefully in order to ensure convergence;
parameterised forms given in Ref.  \onlinecite{seeds} were used as the
initial seeds.

\begin{references}
\bibitem{footnote}
Changes to the inter-layer local 
field factor $G_{12}$ are unlikely to have a significant effect, given 
the weakness of inter-layer corelations demonstrated in Section 
\ref{Theoryres}, and so we can set $G_{12} = 0$.
\end{references}
\end{document}